# Overview of second and third order optical nonlinear processes

A Srinivasa Rao

Department of Physics, Pondicherry University, Puducherry, India-605014

Email: sri.jsp7@gmail.com, asvrao@prl.res.in

**Abstract**

To add the detailed information to the existed knowledge of nonlinear optics field, the nonlinear optical processes of the second and third order have been discussed in their respective susceptibility tensors in the presence of degenerate and non-degenerate optical wave frequencies mixing. This article reports on the all possible ways of wave mixing in the nonlinear interaction of waves in the presence of the medium. The advantage of high peak power pico- and femto-second laser pulses allow us to tune the wavelength of laser beams from UV to infrared region of the electromagnetic spectrum by different harmonic generations. These harmonic generations are described in all aspects and how their efficiency will be affected by other nonlinear processes, those cannot be removed by phase matching condition. The susceptibility terms involved in each nonlinear process are significantly discussed. Also, the probability of occurrence of each nonlinear process is comparatively studied over second and third nonlinear wave mixings.

## 1. Introduction

After the invention of Laser, due to its less etendue and large coherence nature, the study of basic optical phenomena and material characterization are becoming quite easy and straightforward [1-8]. It has created a special branch in the optics called as nonlinear optics [9-11] deals the light-matter interaction at sufficiently high power with reference to the material dielectric constants. The super special characteristics of lasers make the interaction between the light waves in the presence of all kind nonlinear media and the nonlinear optics field become vast [12]. The nonlinear optics was inception when the Second Harmonic Generation (SHG) was discovered by Franken et. al. in 1961[13]. At high intensity, light-matter interaction through the microscopic properties (polarizability, absorption cross-section, and lifetime), led the changes in the macroscopic properties (polarization, absorption, and susceptibility) as a function of interacting intensity [14-17]. In the application point of view, it is necessary to know, how these parameters change with the intensity. These intensity dependent parameters can be measured by different nonlinear techniques [10, 16-19]. From the intensity-dependent macroscopic parameters, one can estimate the susceptibility. The susceptibility tells the material behavior as a function of wavelength and pulse width of the laser. The well studied susceptibility property of the materials led the people to view in all applications perspective. Having developed formalism for the nonlinear optics theory of radiation; we can find it both interesting and profitable in explanation of nonlinear optical phenomena. In this paper, I discussed the susceptibility tensor by

Present address: Photonic Sciences Lab, Physical Research Laboratory, Navarangpura, Ahmedabad 380009, Gujarat, India, email: asvrao@prl.res.in

considering all possible ways of optical waves interaction. The notations used in this article are same as used in the popular nonlinear optics book by Robert W Boyd [20] and all the expressions are given in the Gaussian units. The light-matter interaction of the optical field with its complex conjugation is considered as

$$\mathrm{E}(t) = \mathrm{E}e^{-i\omega t} + \mathrm{E}e^{i\omega t}(\text{c.c}) \tag{1.1}$$

The propagation of electromagnetic wave in the nonlinear medium is

$$\nabla^2 E(r,t) - \frac{n^2}{c^2}\frac{\partial^2 E(r,t)}{\partial t^2} = \frac{4\pi}{c^2}\frac{\partial^2 P(r,t)}{\partial t^2} \tag{1.2}$$

$$\mathrm{P}(t) = \chi^{(1)}\mathrm{E}(t) + \chi^{(2)}\mathrm{E}^2(t) + \chi^{(3)}\mathrm{E}^3(t) + \ldots\ldots\ldots \tag{1.3}$$

The left-hand side term is the generated wave propagation in the medium. The right-hand side inhomogeneous part act as generating and driving source to the wave. Pump and generated waves coupled by this term through nonlinear polarization. Eq. 1.3 gives the nonlinear polarization [21] for different orders. In nonlinear mixing, simultaneously more than one nonlinear processes present in the light-matter interaction. Through phase matching condition, one can enhance the required nonlinear signal by decreasing the other nonlinear processes involved in the nonlinear polarization.The general form of nonlinear polarization can be written as

$$\mathrm{P}_i(\omega_o + \omega_n + \omega_m + \ldots\ldots.) = \sum_{jkl\ldots}\sum_{onm\ldots}\chi^{(\delta)}_{ijkl\ldots}(\omega_o + \omega_n + \omega_m + \ldots.;\omega_o,\omega_n,\omega_m,\ldots..)\mathrm{E}_j(\omega_o)\mathrm{E}_k(\omega_n)\mathrm{E}_l(\omega_m)\ldots.. \tag{1.4}$$

In the susceptibility tensor $\chi_{ijk}(\omega_0+\omega_n+\omega_m+\ldots..;\omega_0,\omega_n,\omega_m, \ldots..)$; $i, j, k,\ldots.$ are representing the Cartesian co-ordinates, and $o, n, m, \ldots.$ are the dummy indices for representation of frequency components. The order of arrangement of frequencies tells the order in which they are interacting with the medium in the time-domain. In Isotropic material, the susceptibility is the scalar quantity and in anisotropic crystals, it becomes tensor matrix. The symmetries in the real material reduce the number of elements in the susceptibility tensor. Depending upon the interaction frequency and frequency of excited energy levels with respect to ground state, the susceptibility follows the different symmetries *viz*: Intrinsic permutation symmetry, Symmetry for lossless media, and Kleinman's Symmetry [20]. In the Kleinman's symmetry, we can write the second and third order susceptibility tensors in the form of D-matrices as

$$d_{ijk} = \frac{1}{2}\chi^{(2)}_{ijk}$$

| | | | | | | |
|---|---|---|---|---|---|---|
| *jk* | 11 | 22 | 33 | 23,32 | 31,13 | 12,21 |
| *l* | 1 | 2 | 3 | 4 | 5 | 6 |

$$d_{il} = \begin{pmatrix} d_{11} & d_{12} & d_{13} & d_{14} & d_{15} & d_{16} \\ d_{16} & d_{22} & d_{23} & d_{24} & d_{14} & d_{12} \\ d_{15} & d_{24} & d_{33} & d_{23} & d_{13} & d_{14} \end{pmatrix} \tag{1.5}$$

$$d_{hijk} = \frac{1}{3}\chi^{(3)}_{hijk}$$

| | | | | | | | | | | |
|---|---|---|---|---|---|---|---|---|---|---|
| *ijk* | 111 | 222 | 333 | 233,323,332 | 322,232,223 | 331,313,133 | 113,131,311 | 122,212,221 | 112,121,211 | 123,132,213,231,312,321 |
| *l* | 1 | 2 | 3 | 4 | 5 | 6 | 7 | 8 | 9 | 10 |

Present address: Photonic Sciences Lab, Physical Research Laboratory, Navarangpura, Ahmedabad 380009, Gujarat, India, email: asvrao@prl.res.in

$$d_{il} = \begin{pmatrix} d_{11} & d_{12} & d_{13} & d_{14} & d_{15} & d_{16} & d_{17} & d_{18} & d_{19} & d_{1,10} \\ d_{19} & d_{22} & d_{23} & d_{24} & d_{25} & d_{14} & d_{1,10} & d_{12} & d_{18} & d_{15} \\ d_{17} & d_{25} & d_{33} & d_{23} & d_{24} & d_{13} & d_{16} & d_{15} & d_{1,10} & d_{14} \end{pmatrix} \tag{1.6}$$

Already exist plenty of books and review articles on the nonlinear optics. The aim of this article is to explore in detail the second and third order nonlinearities in terms of susceptibility tensors. The nonlinear polarization has been discussed under three conditions: resonant excitation, off-resonant excitation, and far off-resonant excitation. In the second section, we concentrated on the second-order nonlinearity. As part of this I have explained about SHG, sum frequency generation (SFG), difference frequency generation (DFG), and optical rectification (OR). The third section dealt with the third order nonlinearity, where we can see the third harmonic generation (THG), sum frequency generation (SFG), Sum with Difference Frequency Generation (SFDFG), Sum with Second Harmonic Generation (SFSHG), and Difference Frequency with Second Harmonic Generation (DFSHG). In the fourth section, I have given a brief description of the nonlinear processes those discussed in the second and third sections. In addition to this, explained the mutual effects in the nonlinear processes, when simultaneously present more than one nonlinear process. The high peak power of pico- and femto-second laser pulses allow us to continuously tune the wavelength in the range of UV-infrared of the electromagnetic spectrum by harmonic generations. To avoid the allowance of the effect of the finite line width, we have considered monochromatic interacting frequencies as pump sources frequencies. For easy to compare the probability of occurring of each nonlinear process, optical amplitudes of incident frequencies are fractionalized in such a way that the sum of the fractions will be one. In anisotropic crystals, their non-symmetric nature makes the scalar susceptibility become tensor form. In our entire discussions, we fractionalized the susceptibility into tensor components so that overall fraction becomes unity.

## 2. Second-order nonlinearity

Second order nonlinearity is the lowest order of nonlinear process in non-centrosymmetric systems. Here two input waves mixe to produce a third wave. Consider two optical fields of frequencies $\omega_1$ and $\omega_2$ with equal amplitudes are interacting in the medium, where the $\omega_1$ is the larger of the two frequencies. The resultant optical field amplitude is

$$\mathrm{E}(t) = \frac{1}{2}\mathrm{E}(\omega_1)e^{-i\omega_1 t} + \frac{1}{2}\mathrm{E}(\omega_2)e^{-i\omega_2 t} + c.c. \tag{2.1}$$

Then the explicit form of second order polarization is

$$\begin{aligned} \mathrm{P}^{(2)}(t) = {} & \frac{1}{4}\chi^{(2)}(2\omega_1;\omega_1,\omega_1)\mathrm{E}^2(\omega_1)e^{-i2\omega_1 t} + \frac{1}{4}\chi^{(2)}(2\omega_2;\omega_2,\omega_2)\mathrm{E}^2(\omega_2)e^{-i2\omega_2 t} \\ & + \frac{1}{4}\left[\chi^{(2)}(\omega_1+\omega_2;\omega_1,\omega_2) + \chi^{(2)}(\omega_2+\omega_1;\omega_2,\omega_1)\right]\mathrm{E}(\omega_1)\mathrm{E}(\omega_2)e^{-i(\omega_1+\omega_2)t} \\ & + \frac{1}{4}\left[\chi^{(2)}(\omega_2-\omega_1;\omega_2,-\omega_1) + \chi^{(2)}(-\omega_1+\omega_2;-\omega_1,\omega_2)\right]\mathrm{E}(\omega_2)\mathrm{E}^*(\omega_1)\left\langle e^{-i\omega_1 t}\middle| e^{-i(\omega_2-\omega_1)t}\right\rangle + c.c. \\ & + \frac{1}{4}\left[\chi^{(2)}(0;\omega_1,-\omega_1) + \chi^{(2)}(0;-\omega_1,\omega_1)\right]\left|\mathrm{E}(\omega_1)\right|^2 + \frac{1}{4}\left[\chi^{(2)}(0;\omega_2,-\omega_2) + \chi^{(2)}(0;-\omega_2,\omega_2)\right]\left|\mathrm{E}(\omega_2)\right|^2 \end{aligned} \tag{2.2}$$

As shown in the Eq. 2.2 the second order polarization contributed by 16 susceptibility tensors (6+6(c.c.) +4=16). First two are SHG, where two photons with the same frequency are interacting within a nonlinear medium to generate new photon with twice the energy of incident photons [22]. The third and fourth terms are corresponding to SFG. Two photons with different frequencies annihilate to produce the third photon. In fifth and sixth terms, two

Present address: Photonic Sciences Lab, Physical Research Laboratory, Navarangpura, Ahmedabad 380009, Gujarat, India, email: asvrao@prl.res.in

photons with different frequencies annihilates in conjugation way to produce the DFG. In these terms, we represented the generated frequency components in the bra-ket notation. Because not as like other terms, the process produces new frequency as well as the frequency which is same as one of the fundamental frequencies. the component in the ket is new difference frequency term and term in the bra is a residual term of frequency in the DFG and it amplifies the corresponding term in the process. Optical Parametric Amplification (OPA) is the best example for DFG [23]. The remaining terms correspond to OR. It was first reported in 1962 [24] and it gives the quasi-static polarization in the non-centrosymmetric materials [25, 26]. The total number of elements in all nonlinear susceptibility tensors is 432{[54+54(c.c)(SHG)]+[54+54(c.c)(SFG)]+[[54+54(c.c)(DFG)]+108(OR)}. In the no phase matching condition, the optical energy is transverse between these terms with including fundamental (incident) frequencies.

*2.1.SHG*

Let a single frequency (ω) laser beam is incident on the crystal with satisfying the phase matching condition $\hat{k}_2=2\hat{k}_1$ then the nonlinearity becomes the combination of SHG and OR. This matching restricts the energy transfer in the ω→2ω direction. The exciting optical amplitude becomes

$$\mathrm{E}(t)=\mathrm{E}(\omega)e^{-i\omega t}+c.c. \tag{2.3}$$

The polarization expression for SHG is

$$\mathrm{P}^{(2)}(t)=\chi^{(2)}(2\omega;\omega,\omega)\mathrm{E}^2(\omega)e^{-i2\omega t}+c.c.+\left[\chi^{(2)}(0;\omega,-\omega)+\chi^{(2)}(0;-\omega,\omega)\right]\mathrm{E}(\omega)|^2 \tag{2.4}$$

The first term in the right side corresponds to SHG and last two are OR. The degeneracy of incident frequencies makes SHG signal independent with respect to the order of interaction i.e., $\chi_{ijk}(2\omega;\omega,\omega)=\chi_{ikj}(2\omega;\omega,\omega)$. As consequence, the number of elements in each individual polarization of Eq. 2.4 are 27 instead of 54. To reduce the occupation of Eq. we have used the elements in the susceptibility tensor in $\chi_{ijk}(\omega,\omega)$ form instead of $\chi_{ijk}(2\omega;\omega,\omega)$ and also, we have folded the susceptibility tensor to reduce its size to the sake of clarity in the expression.

$$\begin{bmatrix} \mathrm{P}_1(2\omega) \\ \\ \mathrm{P}_2(2\omega) \\ \\ \mathrm{P}_3(2\omega) \\ \\ \end{bmatrix}=\frac{1}{27}\begin{bmatrix} \chi^{(2)}_{111}(\omega,\omega) & \chi^{(2)}_{122}(\omega,\omega) & \chi^{(2)}_{133}(\omega,\omega) & \chi^{(2)}_{123}(\omega,\omega) & \chi^{(2)}_{132}(\omega,\omega) \\ \chi^{(2)}_{113}(\omega,\omega) & \chi^{(2)}_{131}(\omega,\omega) & \chi^{(2)}_{112}(\omega,\omega) & \chi^{(2)}_{121}(\omega,\omega) & \\ \chi^{(2)}_{211}(\omega,\omega) & \chi^{(2)}_{222}(\omega,\omega) & \chi^{(2)}_{233}(\omega,\omega) & \chi^{(2)}_{223}(\omega,\omega) & \chi^{(2)}_{232}(\omega,\omega) \\ \chi^{(2)}_{213}(\omega,\omega) & \chi^{(2)}_{231}(\omega,\omega) & \chi^{(2)}_{212}(\omega,\omega) & \chi^{(2)}_{221}(\omega,\omega) & \\ \chi^{(2)}_{311}(\omega,\omega) & \chi^{(2)}_{322}(\omega,\omega) & \chi^{(2)}_{333}(\omega,\omega) & \chi^{(2)}_{323}(\omega,\omega) & \chi^{(2)}_{332}(\omega,\omega) \\ \chi^{(2)}_{313}(\omega,\omega) & \chi^{(2)}_{331}(\omega,\omega) & \chi^{(2)}_{312}(\omega,\omega) & \chi^{(2)}_{321}(\omega,\omega) & \end{bmatrix}\begin{bmatrix} \mathrm{E}_1^2(\omega) \\ \mathrm{E}_2^2(\omega) \\ \mathrm{E}_3^2(\omega) \\ \mathrm{E}_2(\omega)\mathrm{E}_3(\omega) \\ \mathrm{E}_3(\omega)\mathrm{E}_2(\omega) \\ \mathrm{E}_1(\omega)\mathrm{E}_3(\omega) \\ \mathrm{E}_3(\omega)\mathrm{E}_1(\omega) \\ \mathrm{E}_1(\omega)\mathrm{E}_2(\omega) \\ \mathrm{E}_2(\omega)\mathrm{E}_1(\omega) \end{bmatrix} \tag{2.5}$$

$$\begin{bmatrix} \mathrm{P}_1(0) \\ \\ \mathrm{P}_2(0) \\ \\ \mathrm{P}_3(0) \\ \\ \end{bmatrix}=\frac{1}{27}\begin{bmatrix} \chi^{(2)}_{111}(\omega,-\omega) & \chi^{(2)}_{122}(\omega,-\omega) & \chi^{(2)}_{133}(\omega,-\omega) & \chi^{(2)}_{123}(\omega,-\omega) & \chi^{(2)}_{132}(\omega,-\omega) \\ \chi^{(2)}_{113}(\omega,-\omega) & \chi^{(2)}_{131}(\omega,-\omega) & \chi^{(2)}_{112}(\omega,-\omega) & \chi^{(2)}_{121}(\omega,-\omega) & \\ \chi^{(2)}_{211}(\omega,-\omega) & \chi^{(2)}_{222}(\omega,-\omega) & \chi^{(2)}_{233}(\omega,-\omega) & \chi^{(2)}_{223}(\omega,-\omega) & \chi^{(2)}_{232}(\omega,-\omega) \\ \chi^{(2)}_{213}(\omega,-\omega) & \chi^{(2)}_{231}(\omega,-\omega) & \chi^{(2)}_{212}(\omega,-\omega) & \chi^{(2)}_{221}(\omega,-\omega) & \\ \chi^{(2)}_{311}(\omega,-\omega) & \chi^{(2)}_{322}(\omega,-\omega) & \chi^{(2)}_{333}(\omega,-\omega) & \chi^{(2)}_{323}(\omega,-\omega) & \chi^{(2)}_{332}(\omega,-\omega) \\ \chi^{(2)}_{313}(\omega,-\omega) & \chi^{(2)}_{331}(\omega,-\omega) & \chi^{(2)}_{312}(\omega,-\omega) & \chi^{(2)}_{321}(\omega,-\omega) & \end{bmatrix}\begin{bmatrix} \mathrm{E}_1(\omega)\mathrm{E}_1(-\omega) \\ \mathrm{E}_2(\omega)\mathrm{E}_2(-\omega) \\ \mathrm{E}_3(\omega)\mathrm{E}_3(-\omega) \\ \mathrm{E}_2(\omega)\mathrm{E}_3(-\omega) \\ \mathrm{E}_3(\omega)\mathrm{E}_2(-\omega) \\ \mathrm{E}_1(\omega)\mathrm{E}_3(-\omega) \\ \mathrm{E}_3(\omega)\mathrm{E}_1(-\omega) \\ \mathrm{E}_1(\omega)\mathrm{E}_2(-\omega) \\ \mathrm{E}_2(\omega)\mathrm{E}_1(-\omega) \end{bmatrix} \tag{2.6a}$$

Present address: Photonic Sciences Lab, Physical Research Laboratory, Navarangpura, Ahmedabad 380009, Gujarat, India, email: asvrao@prl.res.in

$$\begin{bmatrix} P_1(0) \\ \\ \\ P_2(0) \\ \\ \\ P_3(0) \\ \\ \end{bmatrix} = \frac{1}{27}\begin{bmatrix} \chi^{(2)}_{111}(-\omega,\omega) & \chi^{(2)}_{122}(-\omega,\omega) & \chi^{(2)}_{133}(-\omega,\omega) & \chi^{(2)}_{123}(-\omega,\omega) & \chi^{(2)}_{132}(-\omega,\omega) \\ \chi^{(2)}_{113}(-\omega,\omega) & \chi^{(2)}_{131}(-\omega,\omega) & \chi^{(2)}_{112}(-\omega,\omega) & \chi^{(2)}_{121}(-\omega,\omega) & \\ \\ \chi^{(2)}_{211}(-\omega,\omega) & \chi^{(2)}_{222}(-\omega,\omega) & \chi^{(2)}_{233}(-\omega,\omega) & \chi^{(2)}_{223}(-\omega,\omega) & \chi^{(2)}_{232}(-\omega,\omega) \\ \chi^{(2)}_{213}(-\omega,\omega) & \chi^{(2)}_{231}(-\omega,\omega) & \chi^{(2)}_{212}(-\omega,\omega) & \chi^{(2)}_{221}(-\omega,\omega) & \\ \\ \chi^{(2)}_{311}(-\omega,\omega) & \chi^{(2)}_{322}(-\omega,\omega) & \chi^{(2)}_{333}(-\omega,\omega) & \chi^{(2)}_{323}(-\omega,\omega) & \chi^{(2)}_{332}(-\omega,\omega) \\ \chi^{(2)}_{313}(-\omega,\omega) & \chi^{(2)}_{331}(-\omega,\omega) & \chi^{(2)}_{312}(-\omega,\omega) & \chi^{(2)}_{321}(-\omega,\omega) & \\ \\ \end{bmatrix}\begin{bmatrix} E_1(-\omega)E_1(\omega) \\ E_2(-\omega)E_2(\omega) \\ E_3(-\omega)E_3(\omega) \\ E_2(-\omega)E_3(\omega) \\ E_3(\omega)E_2(\omega) \\ E_1(-\omega)E_3(\omega) \\ E_3(-\omega)E_1(\omega) \\ E_1(-\omega)E_2(\omega) \\ E_2(-\omega)E_1(\omega) \end{bmatrix} \tag{2.6b}$$

Here $j=k$ corresponds to parallel OR and $j\neq k$ is orthogonal OR. Above expressions give what are the possible susceptibility terms present in the SHG and OR. The total number of terms is 108{[27+27(cc)SHG]+54(OR)}. Now consider the case of off-resonant i.e., no one of ω and 2ω is resonant with real energy levels of medium then the real and imaginary parts of susceptibilities become equal and this condition enables us to reduce the susceptibility tensor into Kleinman's D-matrix form which contains 18 independent terms and thus the nonlinear polarization can be written in the D-matrix form as

$$P^{(2)}(t) = 2\left(d^{SHG}_{2\omega}E^2(\omega)e^{-i2\omega t} + d^{OR}_{\omega}\left|E(\omega)\right|^2\right) \tag{2.7}$$

$$\begin{bmatrix} P_1(2\omega) \\ P_2(2\omega) \\ P_3(2\omega) \end{bmatrix} = \frac{1}{27}\begin{pmatrix} d_{11} & d_{12} & d_{13} & d_{14} & d_{15} & d_{16} \\ d_{21} & d_{22} & d_{23} & d_{24} & d_{25} & d_{26} \\ d_{31} & d_{32} & d_{33} & d_{34} & d_{35} & d_{36} \end{pmatrix}\begin{bmatrix} E_1^2(\omega) \\ E_2^2(\omega) \\ E_3^2(\omega) \\ 2E_2(\omega)E_3(\omega) \\ 2E_1(\omega)E_3(\omega) \\ 2E_1(\omega)E_2(\omega) \end{bmatrix} \tag{2.8}$$

$$\begin{bmatrix} P_1(0) \\ P_2(0) \\ P_3(0) \end{bmatrix} = \frac{1}{27}\begin{pmatrix} d_{11} & d_{12} & d_{13} & d_{14} & d_{15} & d_{16} \\ d_{21} & d_{22} & d_{23} & d_{24} & d_{25} & d_{26} \\ d_{31} & d_{32} & d_{33} & d_{34} & d_{35} & d_{36} \end{pmatrix}\begin{bmatrix} E_1(\omega)E_1(-\omega) \\ E_2(\omega)E_2(-\omega) \\ E_3(\omega)E_3(-\omega) \\ 2E_2(\omega)E_3(-\omega) \\ 2E_1(\omega)E_3(-\omega) \\ 2E_1(\omega)E_2(-\omega) \end{bmatrix} \tag{2.9}$$

In this condition, the total number of terms is 32. If the incident optical field strengths are extremely low as compared with first excited state of the material, then the system responds instantaneously to the applied field and by Kleinman's symmetry, we can interchange the frequencies without changing Cartesian co-ordinates. Thus, the D-matrices become

$$d = \frac{1}{27}\begin{pmatrix} d_{11} & d_{12} & d_{13} & d_{14} & d_{15} & d_{16} \\ d_{16} & d_{22} & d_{23} & d_{24} & d_{14} & d_{12} \\ d_{15} & d_{24} & d_{33} & d_{23} & d_{13} & d_{14} \end{pmatrix} \tag{2.10}$$

Therefore the total number of independent terms is 20. Depending upon the symmetries in the anisotropic crystals some of the elements of D-matrix contribution in the nonlinear polarization become zero. To make phase matching in the crystals the incident light must be polarized and should satisfy the refractive index condition $n(\omega)=n(2\omega)$. It can be achieved either by angle tuning or temperature tuning in an anisotropic crystal by using different types of polarization phase matching conditions or by the quasi-phase matching condition in the periodically poled crystals.

Present address: Photonic Sciences Lab, Physical Research Laboratory, Navarangpura, Ahmedabad 380009, Gujarat, India, email: asvrao@prl.res.in

Linear polarization of incident and newly generated beams again reduces the complexity of the susceptibility tensor. In angular tuning method at a precise angle of orientation (θ) of the optical axis of crystal with respect to the incident beam direction, we can optimize the phase matching condition. Here the susceptibility tensor becomes a function of the three spatial co-ordinates. Crystals like lithium niobate [27-32] material by holding the θ at $90^0$ it is possible to phase match by temperature tuning technique. In this scenario, the susceptibility matrix becomes two-dimensionally constrained. Let us suppose beam is propagating in the 3-axis direction then the state of polarization presents in the 12-plan as

$$\begin{bmatrix} P_1(2\omega) \\ P_2(2\omega) \end{bmatrix} = \frac{1}{8}\begin{bmatrix} \chi^{(2)}_{111}(\omega,\omega) & \chi^{(2)}_{122}(\omega,\omega) & \chi^{(2)}_{112}(\omega,\omega) & \chi^{(2)}_{121}(\omega,\omega) \\ \chi^{(2)}_{211}(\omega,\omega) & \chi^{(2)}_{222}(\omega,\omega) & \chi^{(2)}_{212}(\omega,\omega) & \chi^{(2)}_{221}(\omega,\omega) \end{bmatrix}\begin{bmatrix} E_1^2(\omega) \\ E_2^2(\omega) \\ E_1(\omega)E_2(\omega) \\ E_2(\omega)E_1(\omega) \end{bmatrix} \tag{2.11}$$

$$\begin{bmatrix} P_1(0) \\ P_2(0) \end{bmatrix} = \frac{1}{8}\begin{bmatrix} \chi^{(2)}_{111}(\omega,-\omega) & \chi^{(2)}_{122}(\omega,-\omega) & \chi^{(2)}_{112}(\omega,-\omega) & \chi^{(2)}_{121}(\omega,-\omega) \\ \chi^{(2)}_{211}(\omega,-\omega) & \chi^{(2)}_{222}(\omega,-\omega) & \chi^{(2)}_{212}(\omega,-\omega) & \chi^{(2)}_{221}(\omega,-\omega) \end{bmatrix}\begin{bmatrix} E_1(\omega)E_1(-\omega) \\ E_2(\omega)E_2(-\omega) \\ E_1(\omega)E_2(-\omega) \\ E_2(\omega)E_1(-\omega) \end{bmatrix} \tag{2.12a}$$

$$\begin{bmatrix} P_1(0) \\ P_2(0) \end{bmatrix} = \frac{1}{8}\begin{bmatrix} \chi^{(2)}_{111}(-\omega,\omega) & \chi^{(2)}_{122}(-\omega,\omega) & \chi^{(2)}_{112}(-\omega,\omega) & \chi^{(2)}_{121}(-\omega,\omega) \\ \chi^{(2)}_{211}(-\omega,\omega) & \chi^{(2)}_{222}(-\omega,\omega) & \chi^{(2)}_{212}(-\omega,\omega) & \chi^{(2)}_{221}(-\omega,\omega) \end{bmatrix}\begin{bmatrix} E_1(-\omega)E_1(\omega) \\ E_2(-\omega)E_2(\omega) \\ E_1(-\omega)E_2(\omega) \\ E_2(-\omega)E_1(\omega) \end{bmatrix} \tag{2.12b}$$

The total number of terms is 32 {[8+8(cc) SHG]+16(OR)}. At off resonant process the D-matrices are of the form

$$d = \frac{1}{8}\begin{bmatrix} d_{11} & d_{12} & d_{13} \\ d_{21} & d_{22} & d_{23} \end{bmatrix} \tag{2.13}$$

$$\text{Here } 12 = 21 = 3$$

The total number of independent terms is12. In the Kleinman's symmetry number of independent terms reduces to 8 and the D-matrices become

$$d = \frac{1}{8}\begin{bmatrix} d_{11} & d_{12} & d_{13} \\ d_{13} & d_{22} & d_{12} \end{bmatrix} \tag{2.14}$$

The quasi-phase matching is required to use the isotopic materials those have large nonlinearity as compare with anisotropic materials. Generally, the magnitude of diagonal elements is larger than non-diagonal elements in the susceptibility tensor. Through phase matching, we can use the diagonal elements alone to produce the large nonlinear conversion efficiency. Depending upon the selection of the direction only one D-matrix element will be present. Let us consider lithium niobate which shows $d_{33}$ is approximately six times larger than the $d_{31}$. By periodical poles, we can use the$d_{33}$ element and the polarization become

$$P^{(2)}(t) = 2\left(d_{33}^{\text{SHG}} E^2(\omega)e^{-i2\omega t} + d_{33}^{\text{OR}}\left|E(\omega)\right|^2\right) \tag{2.15}$$

*2.2.SFG*

Present address: Photonic Sciences Lab, Physical Research Laboratory, Navarangpura, Ahmedabad 380009, Gujarat, India, email: asvrao@prl.res.in

This process owing to the annihilation of two photons of different frequencies to create the third photon whose frequency is equal to the summation of the annihilation photons frequency. Let $\omega_1$ and $\omega_2$ are added in the nonlinear crystal with phase matching condition $\hat{k}_3=\hat{k}_1+\hat{k}_2$ to produce the third photon with frequency $\omega_1+\omega_2$ then the nonlinear polarization is

$$\begin{aligned} P^{(2)}(t) &= \frac{1}{4}\left[\chi^{(2)}(\omega_1+\omega_2;\omega_1,\omega_2)+\chi^{(2)}(\omega_2+\omega_1;\omega_2,\omega_1)\right]E(\omega_1)E(\omega_2)e^{-i(\omega_1+\omega_2)t}+c.c. \\ &+\frac{1}{4}\left[\chi^{(2)}(0;\omega_1,-\omega_1)+\chi^{(2)}(0;-\omega_1,\omega_1)\right]|E(\omega_1)|^2+\frac{1}{4}\left[\chi^{(2)}(0;\omega_2,-\omega_2)+\chi^{(2)}(0;-\omega_2,\omega_2)\right]|E(\omega_2)|^2 \end{aligned} \tag{2.16}$$

Due to non-degeneracy of fundamental frequencies, the sum frequency contains explicitly two susceptibility tensors of each contains 27 elements and OR can be explicitly written as sum of two terms for each fundamental frequency. The OR direction of each frequency is controlled by their optical field's polarization direction. Susceptibility tensors in the SFG and OR are in their explicit form are

$$\begin{bmatrix} P_1(\omega_1+\omega_2) \\ P_2(\omega_1+\omega_2) \\ P_3(\omega_1+\omega_2) \end{bmatrix} = \frac{1}{27}\begin{bmatrix} \chi^{(2)}_{111}(\omega_1,\omega_2) & \chi^{(2)}_{122}(\omega_1,\omega_2) & \chi^{(2)}_{133}(\omega_1,\omega_2) & \chi^{(2)}_{123}(\omega_1,\omega_2) & \chi^{(2)}_{132}(\omega_1,\omega_2) & \chi^{(2)}_{113}(\omega_1,\omega_2) & \chi^{(2)}_{131}(\omega_1,\omega_2) & \chi^{(2)}_{112}(\omega_1,\omega_2) & \chi^{(2)}_{121}(\omega_1,\omega_2) \\ \chi^{(2)}_{211}(\omega_1,\omega_2) & \chi^{(2)}_{222}(\omega_1,\omega_2) & \chi^{(2)}_{233}(\omega_1,\omega_2) & \chi^{(2)}_{223}(\omega_1,\omega_2) & \chi^{(2)}_{232}(\omega_1,\omega_2) & \chi^{(2)}_{213}(\omega_1,\omega_2) & \chi^{(2)}_{231}(\omega_1,\omega_2) & \chi^{(2)}_{212}(\omega_1,\omega_2) & \chi^{(2)}_{221}(\omega_1,\omega_2) \\ \chi^{(2)}_{311}(\omega_1,\omega_2) & \chi^{(2)}_{322}(\omega_1,\omega_2) & \chi^{(2)}_{333}(\omega_1,\omega_2) & \chi^{(2)}_{323}(\omega_1,\omega_2) & \chi^{(2)}_{332}(\omega_1,\omega_2) & \chi^{(2)}_{313}(\omega_1,\omega_2) & \chi^{(2)}_{331}(\omega_1,\omega_2) & \chi^{(2)}_{312}(\omega_1,\omega_2) & \chi^{(2)}_{321}(\omega_1,\omega_2) \end{bmatrix}\begin{bmatrix} E_1(\omega_1)E_1(\omega_2) \\ E_2(\omega_1)E_2(\omega_2) \\ E_3(\omega_1)E_3(\omega_2) \\ E_2(\omega_1)E_3(\omega_2) \\ E_3(\omega_1)E_2(\omega_2) \\ E_1(\omega_1)E_3(\omega_2) \\ E_3(\omega_1)E_1(\omega_2) \\ E_1(\omega_1)E_2(\omega_2) \\ E_2(\omega_1)E_1(\omega_2) \end{bmatrix} \tag{2.17a}$$

$$\begin{bmatrix} P_1(\omega_2+\omega_1) \\ P_2(\omega_2+\omega_1) \\ P_3(\omega_2+\omega_1) \end{bmatrix} = \frac{1}{27}\begin{bmatrix} \chi^{(2)}_{111}(\omega_2,\omega_1) & \chi^{(2)}_{122}(\omega_2,\omega_1) & \chi^{(2)}_{133}(\omega_2,\omega_1) & \chi^{(2)}_{123}(\omega_2,\omega_1) & \chi^{(2)}_{132}(\omega_2,\omega_1) & \chi^{(2)}_{113}(\omega_2,\omega_1) & \chi^{(2)}_{131}(\omega_2,\omega_1) & \chi^{(2)}_{112}(\omega_2,\omega_1) & \chi^{(2)}_{121}(\omega_2,\omega_1) \\ \chi^{(2)}_{211}(\omega_2,\omega_1) & \chi^{(2)}_{222}(\omega_2,\omega_1) & \chi^{(2)}_{233}(\omega_2,\omega_1) & \chi^{(2)}_{223}(\omega_2,\omega_1) & \chi^{(2)}_{232}(\omega_2,\omega_1) & \chi^{(2)}_{213}(\omega_2,\omega_1) & \chi^{(2)}_{231}(\omega_2,\omega_1) & \chi^{(2)}_{212}(\omega_2,\omega_1) & \chi^{(2)}_{221}(\omega_2,\omega_1) \\ \chi^{(2)}_{311}(\omega_2,\omega_1) & \chi^{(2)}_{322}(\omega_2,\omega_1) & \chi^{(2)}_{333}(\omega_2,\omega_1) & \chi^{(2)}_{323}(\omega_2,\omega_1) & \chi^{(2)}_{332}(\omega_2,\omega_1) & \chi^{(2)}_{313}(\omega_2,\omega_1) & \chi^{(2)}_{331}(\omega_2,\omega_1) & \chi^{(2)}_{312}(\omega_2,\omega_1) & \chi^{(2)}_{321}(\omega_2,\omega_1) \end{bmatrix}\begin{bmatrix} E_1(\omega_2)E_1(\omega_1) \\ E_2(\omega_2)E_2(\omega_1) \\ E_3(\omega_2)E_3(\omega_1) \\ E_2(\omega_2)E_3(\omega_1) \\ E_3(\omega_2)E_2(\omega_1) \\ E_1(\omega_2)E_3(\omega_1) \\ E_3(\omega_2)E_1(\omega_1) \\ E_1(\omega_2)E_2(\omega_1) \\ E_2(\omega_2)E_1(\omega_1) \end{bmatrix} \tag{2.17b}$$

$$\begin{bmatrix} P_1(0) \\ P_2(0) \\ P_3(0) \end{bmatrix} = \frac{1}{27}\begin{bmatrix} \chi^{(2)}_{111}(\omega_1,-\omega_1) & \chi^{(2)}_{122}(\omega_1,-\omega_1) & \chi^{(2)}_{133}(\omega_1,-\omega_1) & \chi^{(2)}_{123}(\omega_1,-\omega_1) & \chi^{(2)}_{132}(\omega_1,-\omega_1) & \chi^{(2)}_{113}(\omega_1,-\omega_1) & \chi^{(2)}_{131}(\omega_1,-\omega_1) & \chi^{(2)}_{112}(\omega_1,-\omega_1) & \chi^{(2)}_{121}(\omega_1,-\omega_1) \\ \chi^{(2)}_{211}(\omega_1,-\omega_1) & \chi^{(2)}_{222}(\omega_1,-\omega_1) & \chi^{(2)}_{233}(\omega_1,-\omega_1) & \chi^{(2)}_{223}(\omega_1,-\omega_1) & \chi^{(2)}_{232}(\omega_1,-\omega_1) & \chi^{(2)}_{213}(\omega_1,-\omega_1) & \chi^{(2)}_{231}(\omega_1,-\omega_1) & \chi^{(2)}_{212}(\omega_1,-\omega_1) & \chi^{(2)}_{221}(\omega_1,-\omega_1) \\ \chi^{(2)}_{311}(\omega_1,-\omega_1) & \chi^{(2)}_{322}(\omega_1,-\omega_1) & \chi^{(2)}_{333}(\omega_1,-\omega_1) & \chi^{(2)}_{323}(\omega_1,-\omega_1) & \chi^{(2)}_{332}(\omega_1,-\omega_1) & \chi^{(2)}_{313}(\omega_1,-\omega_1) & \chi^{(2)}_{331}(\omega_1,-\omega_1) & \chi^{(2)}_{312}(\omega_1,-\omega_1) & \chi^{(2)}_{321}(\omega_1,-\omega_1) \end{bmatrix}\begin{bmatrix} E_1(\omega_1)E_1(-\omega_1) \\ E_2(\omega_1)E_2(-\omega_1) \\ E_3(\omega_1)E_3(-\omega_1) \\ E_2(\omega_1)E_3(-\omega_1) \\ E_3(\omega_1)E_2(-\omega_1) \\ E_1(\omega_1)E_3(-\omega_1) \\ E_3(\omega_1)E_1(-\omega_1) \\ E_1(\omega_1)E_2(-\omega_1) \\ E_2(\omega_1)E_1(-\omega_1) \end{bmatrix} \tag{2.18a}$$

$$\begin{bmatrix} P_1(0) \\ P_2(0) \\ P_3(0) \end{bmatrix} = \frac{1}{27}\begin{bmatrix} \chi^{(2)}_{111}(-\omega_1,\omega_1) & \chi^{(2)}_{122}(-\omega_1,\omega_1) & \chi^{(2)}_{133}(-\omega_1,\omega_1) & \chi^{(2)}_{123}(-\omega_1,\omega_1) & \chi^{(2)}_{132}(-\omega_1,\omega_1) & \chi^{(2)}_{113}(-\omega_1,\omega_1) & \chi^{(2)}_{131}(-\omega_1,\omega_1) & \chi^{(2)}_{112}(-\omega_1,\omega_1) & \chi^{(2)}_{121}(-\omega_1,\omega_1) \\ \chi^{(2)}_{211}(-\omega_1,\omega_1) & \chi^{(2)}_{222}(-\omega_1,\omega_1) & \chi^{(2)}_{233}(-\omega_1,\omega_1) & \chi^{(2)}_{223}(-\omega_1,\omega_1) & \chi^{(2)}_{232}(-\omega_1,\omega_1) & \chi^{(2)}_{213}(-\omega_1,\omega_1) & \chi^{(2)}_{231}(-\omega_1,\omega_1) & \chi^{(2)}_{212}(-\omega_1,\omega_1) & \chi^{(2)}_{221}(-\omega_1,\omega_1) \\ \chi^{(2)}_{311}(-\omega_1,\omega_1) & \chi^{(2)}_{322}(-\omega_1,\omega_1) & \chi^{(2)}_{333}(-\omega_1,\omega_1) & \chi^{(2)}_{323}(-\omega_1,\omega_1) & \chi^{(2)}_{332}(-\omega_1,\omega_1) & \chi^{(2)}_{313}(-\omega_1,\omega_1) & \chi^{(2)}_{331}(-\omega_1,\omega_1) & \chi^{(2)}_{312}(-\omega_1,\omega_1) & \chi^{(2)}_{321}(-\omega_1,\omega_1) \end{bmatrix}\begin{bmatrix} E_1(-\omega_1)E_1(\omega_1) \\ E_2(-\omega_1)E_2(\omega_1) \\ E_3(-\omega_1)E_3(\omega_1) \\ E_2(-\omega_1)E_3(\omega_1) \\ E_3(-\omega_1)E_2(\omega_1) \\ E_1(\omega-\omega_1)E_3(\omega_1) \\ E_3(-\omega_1)E_1(\omega_1) \\ E_1(-\omega_1)E_2(\omega_1) \\ E_2(-\omega_1)E_1(\omega_1) \end{bmatrix} \tag{2.18b}$$

Present address: Photonic Sciences Lab, Physical Research Laboratory, Navarangpura, Ahmedabad 380009, Gujarat, India, email: asvrao@prl.res.in

$$\begin{bmatrix} P_1(0) \\ \\ \\ P_2(0) \\ \\ \\ P_3(0) \\ \\ \end{bmatrix} = \frac{1}{27}\begin{bmatrix} \chi^{(2)}_{111}(\omega_2,-\omega_2) & \chi^{(2)}_{122}(\omega_2,-\omega_2) & \chi^{(2)}_{133}(\omega_2,-\omega_2) & \chi^{(2)}_{123}(\omega_2,-\omega_2) & \chi^{(2)}_{132}(\omega_2,-\omega_2) \\ \chi^{(2)}_{113}(\omega_2,-\omega_2) & \chi^{(2)}_{131}(\omega_2,-\omega_2) & \chi^{(2)}_{112}(\omega_2,-\omega_2) & \chi^{(2)}_{121}(\omega_2,-\omega_2) & \\ \\ \chi^{(2)}_{211}(\omega_2,-\omega_2) & \chi^{(2)}_{222}(\omega_2,-\omega_2) & \chi^{(2)}_{233}(\omega_2,-\omega_2) & \chi^{(2)}_{223}(\omega_2,-\omega_2) & \chi^{(2)}_{232}(\omega_2,-\omega_2) \\ \chi^{(2)}_{213}(\omega_2,-\omega_2) & \chi^{(2)}_{231}(\omega_2,-\omega_2) & \chi^{(2)}_{212}(\omega_2,-\omega_2) & \chi^{(2)}_{221}(\omega_2,-\omega_2) & \\ \\ \chi^{(2)}_{311}(\omega_2,-\omega_2) & \chi^{(2)}_{322}(\omega_2,-\omega_2) & \chi^{(2)}_{333}(\omega_2,-\omega_2) & \chi^{(2)}_{323}(\omega_2,-\omega_2) & \chi^{(2)}_{332}(\omega_2,-\omega_2) \\ \chi^{(2)}_{313}(\omega_2,-\omega_2) & \chi^{(2)}_{331}(\omega_2,-\omega_2) & \chi^{(2)}_{312}(\omega_2,-\omega_2) & \chi^{(2)}_{321}(\omega_2,-\omega_2) & \\ \\ \end{bmatrix}\begin{bmatrix} E_1(\omega_2)E_1(-\omega_2) \\ E_2(\omega_2)E_2(-\omega_2) \\ E_3(\omega_2)E_3(-\omega_2) \\ E_2(\omega_2)E_3(-\omega_2) \\ E_3(\omega_2)E_2(-\omega_2) \\ E_1(\omega_2)E_3(-\omega_2) \\ E_3(\omega_2)E_1(-\omega_2) \\ E_1(\omega_2)E_2(-\omega_2) \\ E_2(\omega_2)E_1(-\omega_2) \end{bmatrix} \tag{2.18c}$$

$$\begin{bmatrix} P_1(0) \\ \\ \\ P_2(0) \\ \\ \\ P_3(0) \\ \\ \end{bmatrix} = \frac{1}{27}\begin{bmatrix} \chi^{(2)}_{111}(-\omega_2,\omega_2) & \chi^{(2)}_{122}(-\omega_2,\omega_2) & \chi^{(2)}_{133}(-\omega_2,\omega_2) & \chi^{(2)}_{123}(-\omega_2,\omega_2) & \chi^{(2)}_{132}(-\omega_2,\omega_2) \\ \chi^{(2)}_{113}(-\omega_2,\omega_2) & \chi^{(2)}_{131}(-\omega_2,\omega_2) & \chi^{(2)}_{112}(-\omega_2,\omega_2) & \chi^{(2)}_{121}(-\omega_2,\omega_2) & \\ \\ \chi^{(2)}_{211}(-\omega_2,\omega_2) & \chi^{(2)}_{222}(-\omega_2,\omega_2) & \chi^{(2)}_{233}(-\omega_2,\omega_2) & \chi^{(2)}_{223}(-\omega_2,\omega_2) & \chi^{(2)}_{232}(-\omega_2,\omega_2) \\ \chi^{(2)}_{213}(-\omega_2,\omega_2) & \chi^{(2)}_{231}(-\omega_2,\omega_2) & \chi^{(2)}_{212}(-\omega_2,\omega_2) & \chi^{(2)}_{221}(-\omega_2,\omega_2) & \\ \\ \chi^{(2)}_{311}(-\omega_2,\omega_2) & \chi^{(2)}_{322}(-\omega_2,\omega_2) & \chi^{(2)}_{333}(-\omega_2,\omega_2) & \chi^{(2)}_{323}(-\omega_2,\omega_2) & \chi^{(2)}_{332}(-\omega_2,\omega_2) \\ \chi^{(2)}_{313}(-\omega_2,\omega_2) & \chi^{(2)}_{331}(-\omega_2,\omega_2) & \chi^{(2)}_{312}(-\omega_2,\omega_2) & \chi^{(2)}_{321}(-\omega_2,\omega_2) & \\ \\ \end{bmatrix}\begin{bmatrix} E_1(-\omega_2)E_1(\omega_2) \\ E_2(-\omega_2)E_2(\omega_2) \\ E_3(-\omega_2)E_3(\omega_2) \\ E_2(-\omega_2)E_3(\omega_2) \\ E_3(-\omega_2)E_2(\omega_2) \\ E_1(-\omega_2)E_3(\omega_2) \\ E_3(-\omega_2)E_1(\omega_2) \\ E_1(-\omega_2)E_2(\omega_2) \\ E_2(-\omega_2)E_1(\omega_2) \end{bmatrix} \tag{2.19d}$$

Total number of terms is 216 {[54+54(cc) (SFG)]+[108(OR)]}. In the off-resonance case the polarization is

$$P^{(2)}(t) = d^{SFG}_{\omega_1+\omega_2} E(\omega_1)E(\omega_2)e^{-i(\omega_1+\omega_2)t} + d^{OR}_{\omega_1}|E(\omega_1)|^2 + d^{OR}_{\omega_2}|E(\omega_2)|^2 \tag{2.20}$$

Here we have not given D-matrix representation for SFG because they are same as discussed in the SHG. Total number of independent terms is 54. Under Klienman's symmetry the D-matrices reduces with 30 independent variables.

*2.3.DFG*

To make the energy transfer from fundamental frequencies ($\omega_1$ and $\omega_2$) to $\omega_2$-$\omega_1$ it is required to satisfy the phase matching condition $\hat{k}_3=\hat{k}_2-\hat{k}_1$. In this process, not as like previous two processes, two photons interact and three photons come as output. As shown in the first term of Eq. 2.21, two $\omega_1$ and $\omega_2$ photons interact and then produce three photons $\omega_1$, $\omega_1$, and $\omega_2$-$\omega_1$ due to stimulation emission of $\omega_1$ by $\omega_1$. By this process, we can amplify the intensity $\omega_1$ beam and is called as parametric amplification [33-36]. Pumping narrow difference frequencies as pump sources ($\omega_2$-$\omega_1$= $\Delta\omega$ present in the terahertz frequency range), researchers are able to generate terahertz frequencies in BNA and DAST crystals [37-40]. Further, we can generate the terahertz radiation through OR [41].In this process beam corresponds to $\omega_1$ frequency is called as signal and $\omega_2$ beam called as either pump or probe. In this process, $\omega_1$ and $\omega_2$-$\omega_1$beams intensity increase with decreasing $\omega_2$ intensity.

$$\begin{aligned} P^{(2)}(t) = {} & \frac{1}{4}\left[\chi^{(2)}(\omega_2-\omega_1;\omega_2,\omega_1) + \chi^{(2)}(-\omega_1+\omega_2;-\omega_1,\omega_2)\right]E(\omega_2)E^*(\omega_1)\left\langle e^{-i\omega_1 t}\,\middle|\, e^{-i(\omega_2-\omega_1)t}\right\rangle + c.c. \\ & + \frac{1}{4}\left[\chi^{(2)}(0;\omega_1,-\omega_1) + \chi^{(2)}(0;-\omega_1,\omega_1)\right]|E(\omega_1)|^2 + \frac{1}{4}\left[\chi^{(2)}(0;\omega_2,-\omega_2) + \chi^{(2)}(0;-\omega_2,\omega_2)\right]|E(\omega_2)|^2 \end{aligned} \tag{2.21}$$

Polarizations in tensor forms are

Present address: Photonic Sciences Lab, Physical Research Laboratory, Navarangpura, Ahmedabad 380009, Gujarat, India, email: asvrao@prl.res.in

$$
\begin{bmatrix} P_1(\omega_2-\omega_1) \\ \\ P_2(\omega_2-\omega_1) \\ \\ P_3(\omega_2-\omega_1) \\ \\ \end{bmatrix}
=\frac{1}{27}
\begin{bmatrix}
\chi^{(2)}_{111}(\omega_2,-\omega_1) & \chi^{(2)}_{122}(\omega_2,-\omega_1) & \chi^{(2)}_{133}(\omega_2,-\omega_1) & \chi^{(2)}_{123}(\omega_2,-\omega_1) & \chi^{(2)}_{132}(\omega_2,-\omega_1) \\
\chi^{(2)}_{113}(\omega_2,-\omega_1) & \chi^{(2)}_{131}(\omega_2,-\omega_1) & \chi^{(2)}_{112}(\omega_2,-\omega_1) & \chi^{(2)}_{121}(\omega_2,-\omega_1) & \\
\chi^{(2)}_{211}(\omega_2,-\omega_1) & \chi^{(2)}_{222}(\omega_2,-\omega_1) & \chi^{(2)}_{233}(\omega_2,-\omega_1) & \chi^{(2)}_{223}(\omega_2,-\omega_1) & \chi^{(2)}_{232}(\omega_2,-\omega_1) \\
\chi^{(2)}_{213}(\omega_2,-\omega_1) & \chi^{(2)}_{231}(\omega_2,-\omega_1) & \chi^{(2)}_{212}(\omega_2,-\omega_1) & \chi^{(2)}_{221}(\omega_2,-\omega_1) & \\
\chi^{(2)}_{311}(\omega_2,-\omega_1) & \chi^{(2)}_{322}(\omega_2,-\omega_1) & \chi^{(2)}_{333}(\omega_2,-\omega_1) & \chi^{(2)}_{323}(\omega_2,-\omega_1) & \chi^{(2)}_{332}(\omega_2,-\omega_1) \\
\chi^{(2)}_{313}(\omega_2,-\omega_1) & \chi^{(2)}_{331}(\omega_2,-\omega_1) & \chi^{(2)}_{312}(\omega_2,-\omega_1) & \chi^{(2)}_{321}(\omega_2,-\omega_1) &
\end{bmatrix}
\begin{bmatrix}
E_1(\omega_2)E_1(-\omega_1) \\ E_2(\omega_2)E_2(-\omega_1) \\ E_3(\omega_2)E_3(-\omega_1) \\ E_2(\omega_2)E_3(-\omega_1) \\ E_3(\omega_2)E_2(-\omega_1) \\ E_1(\omega_2)E_3(-\omega_1) \\ E_3(\omega_2)E_1(-\omega_1) \\ E_1(\omega_2)E_2(-\omega_1) \\ E_2(\omega_2)E_1(-\omega_1)
\end{bmatrix}
\quad (2.22a)
$$

$$
\begin{bmatrix} P_1(-\omega_1+\omega_2) \\ \\ P_2(-\omega_1+\omega_2) \\ \\ P_3(-\omega_1+\omega_2) \\ \\ \end{bmatrix}
=\frac{1}{27}
\begin{bmatrix}
\chi^{(2)}_{111}(-\omega_1,\omega_2) & \chi^{(2)}_{122}(-\omega_1,\omega_2) & \chi^{(2)}_{133}(-\omega_1,\omega_2) & \chi^{(2)}_{123}(-\omega_1,\omega_2) & \chi^{(2)}_{132}(-\omega_1,\omega_2) \\
\chi^{(2)}_{113}(-\omega_1,\omega_2) & \chi^{(2)}_{131}(-\omega_1,\omega_2) & \chi^{(2)}_{112}(-\omega_1,\omega_2) & \chi^{(2)}_{121}(-\omega_1,\omega_2) & \\
\chi^{(2)}_{211}(-\omega_1,\omega_2) & \chi^{(2)}_{222}(-\omega_1,\omega_2) & \chi^{(2)}_{233}(-\omega_1,\omega_2) & \chi^{(2)}_{223}(-\omega_1,\omega_2) & \chi^{(2)}_{232}(-\omega_1,\omega_2) \\
\chi^{(2)}_{213}(-\omega_1,\omega_2) & \chi^{(2)}_{231}(-\omega_1,\omega_2) & \chi^{(2)}_{212}(-\omega_1,\omega_2) & \chi^{(2)}_{221}(-\omega_1,\omega_2) & \\
\chi^{(2)}_{311}(-\omega_1,\omega_2) & \chi^{(2)}_{322}(-\omega_1,\omega_2) & \chi^{(2)}_{333}(-\omega_1,\omega_2) & \chi^{(2)}_{323}(-\omega_1,\omega_2) & \chi^{(2)}_{332}(-\omega_1,\omega_2) \\
\chi^{(2)}_{313}(-\omega_1,\omega_2) & \chi^{(2)}_{331}(-\omega_1,\omega_2) & \chi^{(2)}_{312}(-\omega_1,\omega_2) & \chi^{(2)}_{321}(-\omega_1,\omega_2) &
\end{bmatrix}
\begin{bmatrix}
E_1(-\omega_1)E_1(\omega_2) \\ E_2(-\omega_1)E_2(\omega_2) \\ E_3(-\omega_1)E_3(\omega_2) \\ E_2(-\omega_1)E_3(\omega_2) \\ E_3(-\omega_1)E_2(\omega_2) \\ E_1(-\omega_1)E_3(\omega_2) \\ E_3(-\omega_1)E_1(\omega_2) \\ E_1(-\omega_1)E_2(\omega_2) \\ E_2(-\omega_1)E_1(\omega_2)
\end{bmatrix}
\quad (2.22b)
$$

We have not given OR tensors because they are same as discussed in the SFG. Total number of terms is 216 {[54+54(cc) (DFG)]+[108(OR)]}. In the off-resonance case the polarization is

$$P^{(2)}(t)=d^{\mathrm{DFG}}_{\omega_2-\omega_1}E^*(\omega_1)E(\omega_2)e^{-i(\omega_2-\omega_1)t}+d^{\mathrm{OR}}_{\omega_1}\left|E(\omega_1)\right|^2+d^{\mathrm{OR}}_{\omega_2}\left|E(\omega_2)\right|^2 \quad (2.23)$$

From Eq. 2.6, 2.20 and 2.23, the probability of generating SHG is double than the SFG and DFG signal. In the general expression of second-order nonlinearity (Eq. 2.2), the first four terms satisfy the phase matching condition in the SHG and in case of SFG and DFG, due to non-degeneracy in the fundamental frequencies only corresponding terms satisfy the phase matching. Thus, it requires large intensity to produce SFG and DFG as compared with SHG. Mathematically the probability of producing SFG and DFG are same but physically to produce the DFG, we need a large number of photons as compare with SFG due to the requirement of stimulated emission. So, the probability of producing SFG is more efficient than DFG.

**3. Third-order nonlinearity**

Third-order nonlinear optical phenomenon is a four-wave mixing process. At a given time three waves mix and generate fourth wave. Consider three waves at frequencies $\omega_1$, $\omega_2$, and $\omega_3$ participating in the third order nonlinear process then the optical amplitude of nonlinear process is

$$E(t)=\frac{1}{3}E(\omega_1)e^{-i\omega_1 t}+\frac{1}{3}E(\omega_2)e^{-i\omega_2 t}+\frac{1}{3}E(\omega_3)e^{-i\omega_3 t}+c.c. \quad (3.1)$$

The explicit form of third-order nonlinear polarization is given by

$$P^{(3)}(t)=\frac{1}{27}\left\{\chi^{(3)}(3\omega_1;\omega_1,\omega_1,\omega_1)E^3(\omega_1)e^{-i3\omega_1 t}+\chi^{(3)}(3\omega_2;\omega_2,\omega_2,\omega_2)E^3(\omega_2)e^{-i3\omega_2 t}+\chi^{(3)}(3\omega_3;\omega_3,\omega_3,\omega_3)E^3(\omega_3)e^{-i3\omega_3 t}\right\}$$

Present address: Photonic Sciences Lab, Physical Research Laboratory, Navarangpura, Ahmedabad 380009, Gujarat, India, email: asvrao@prl.res.in

$$
+\frac{1}{27}\left\{\begin{array}{l}
\left[\chi^{(3)}(2\omega_1+\omega_2;\omega_1,\omega_1,\omega_2)+\chi^{(3)}(\omega_1+\omega_2+\omega_1;\omega_1,\omega_2,\omega_1)+\chi^{(3)}(\omega_2+2\omega_1;\omega_2,\omega_1,\omega_1)\right]\mathrm{E}^2(\omega_1)\mathrm{E}(\omega_2)e^{-i(2\omega_1+\omega_2)t} \\
+\left[\chi^{(3)}(2\omega_1+\omega_3;\omega_1,\omega_1,\omega_3)+\chi^{(3)}(\omega_1+\omega_3+\omega_1;\omega_1,\omega_3,\omega_1)+\chi^{(3)}(\omega_3+2\omega_1;\omega_3,\omega_1,\omega_1)\right]\mathrm{E}^2(\omega_1)\mathrm{E}(\omega_3)e^{-i(2\omega_1+\omega_3)t} \\
+\left[\chi^{(3)}(\omega_1+2\omega_2;\omega_1,\omega_2,\omega_2)+\chi^{(3)}(\omega_2+\omega_1+\omega_2;\omega_2,\omega_1,\omega_2)+\chi^{(3)}(2\omega_2+\omega_1;\omega_2,\omega_2,\omega_1)\right]\mathrm{E}(\omega_1)\mathrm{E}^2(\omega_2)e^{-i(\omega_1+2\omega_2)t} \\
+\left[\chi^{(3)}(2\omega_2+\omega_3;\omega_2,\omega_2,\omega_3)+\chi^{(3)}(\omega_2+\omega_3+\omega_2;\omega_2,\omega_3,\omega_2)+\chi^{(3)}(\omega_3+2\omega_2;\omega_3,\omega_2,\omega_2)\right]\mathrm{E}^2(\omega_2)\mathrm{E}(\omega_3)e^{-i(2\omega_2+\omega_3)t} \\
+\left[\chi^{(3)}(2\omega_3+\omega_2;\omega_3,\omega_3,\omega_2)+\chi^{(3)}(\omega_3+\omega_2+\omega_3;\omega_3,\omega_2,\omega_3)+\chi^{(3)}(\omega_2+2\omega_3;\omega_2,\omega_3,\omega_3)\right]\mathrm{E}^2(\omega_3)\mathrm{E}(\omega_2)e^{-i(2\omega_3+\omega_2)t} \\
+\left[\chi^{(3)}(2\omega_3+\omega_1;\omega_3,\omega_3,\omega_1)+\chi^{(3)}(\omega_3+\omega_1+\omega_3;\omega_3,\omega_1,\omega_3)+\chi^{(3)}(\omega_1+2\omega_3;\omega_1,\omega_3,\omega_3)\right]\mathrm{E}^2(\omega_3)\mathrm{E}(\omega_1)e^{-i(2\omega_3+\omega_1)t}
\end{array}\right\}
$$

$$
+\frac{1}{27}\left\{\begin{array}{l}
\chi^{(3)}(\omega_1+\omega_2+\omega_3;\omega_1,\omega_2,\omega_3)+\chi^{(3)}(\omega_1+\omega_3+\omega_2;\omega_1,\omega_3,\omega_2) \\
+\chi^{(3)}(\omega_2+\omega_1+\omega_3;\omega_2,\omega_1,\omega_3)+\chi^{(3)}(\omega_2+\omega_3+\omega_1;\omega_2,\omega_3,\omega_1) \\
+\chi^{(3)}(\omega_3+\omega_1+\omega_2;\omega_3,\omega_1,\omega_2)+\chi^{(3)}(\omega_3+\omega_2+\omega_1;,\omega_3,\omega_2,\omega_1)
\end{array}\right\}\mathrm{E}(\omega_1)\mathrm{E}(\omega_2)\mathrm{E}(\omega_3)e^{-i(\omega_1+\omega_2+\omega_3)t}
$$

$$
+\frac{1}{27}\left\{\begin{array}{l}
\left[\chi^{(3)}(2\omega_1-\omega_2;\omega_1,\omega_1,-\omega_2)+\chi^{(3)}(\omega_1-\omega_2+\omega_1;\omega_1,-\omega_2,\omega_1)+\chi^{(3)}(-\omega_2+2\omega_1;-\omega_2,\omega_1,\omega_1)\right]\mathrm{E}^2(\omega_1)\mathrm{E}^*(\omega_2)\left\langle e^{-i\omega_2}\middle| e^{-i(2\omega_1-\omega_2)t}\right\rangle \\
+\left[\chi^{(3)}(2\omega_1-\omega_3;\omega_1,\omega_1,-\omega_3)+\chi^{(3)}(\omega_1-\omega_3+\omega_1;\omega_1,-\omega_3,\omega_1)+\chi^{(3)}(-\omega_3+2\omega_1;-\omega_3,\omega_1,\omega_1)\right]\mathrm{E}^2(\omega_1)\mathrm{E}^*(\omega_3)\left\langle e^{-i\omega_3 t}\middle| e^{-i(2\omega_1-\omega_3)t}\right\rangle \\
+\left[\chi^{(3)}(-\omega_1+2\omega_2;-\omega_1,\omega_2,\omega_2)+\chi^{(3)}(\omega_2-\omega_1+\omega_2;\omega_2,-\omega_1,\omega_2)+\chi^{(3)}(2\omega_2-\omega_1;\omega_2,\omega_2,-\omega_1)\right]\mathrm{E}^*(\omega_1)\mathrm{E}^2(\omega_2)\left\langle e^{-i\omega_1 t}\middle| e^{-i(-\omega_1+2\omega_2)t}\right\rangle \\
+\left[\chi^{(3)}(2\omega_2-\omega_3;\omega_2,\omega_2,-\omega_3)+\chi^{(3)}(\omega_2-\omega_3+\omega_2;\omega_2,-\omega_3,\omega_2)+\chi^{(3)}(-\omega_3+2\omega_2;-\omega_3,\omega_2,\omega_2)\right]\mathrm{E}^2(\omega_2)\mathrm{E}^*(\omega_3)\left\langle e^{-i\omega_3 t}\middle| e^{-i(2\omega_2-\omega_3)t}\right\rangle \\
+\left[\chi^{(3)}(2\omega_3-\omega_2;\omega_3,\omega_3,-\omega_2)+\chi^{(3)}(\omega_3-\omega_2+\omega_3;\omega_3,-\omega_2,\omega_3)+\chi^{(3)}(-\omega_2+2\omega_3;-\omega_2,\omega_3,\omega_3)\right]\mathrm{E}^2(\omega_3)\mathrm{E}^*(\omega_2)\left\langle e^{-i\omega_3 t}\middle| e^{-i(2\omega_2-\omega_3)t}\right\rangle \\
+\left[\chi^{(3)}(2\omega_3-\omega_1;\omega_3,\omega_3,-\omega_1)+\chi^{(3)}(\omega_3-\omega_1+\omega_3;\omega_3,-\omega_1,\omega_3)+\chi^{(3)}(-\omega_1+2\omega_3;-\omega_1,\omega_3,\omega_3)\right]\mathrm{E}^2(\omega_3)\mathrm{E}^*(\omega_1)\left\langle e^{-i\omega_1 t}\middle| e^{-i(2\omega_3-\omega_1)t}\right\rangle
\end{array}\right\}
$$

$$
+\frac{1}{27}\left\{\begin{array}{l}
\left[\begin{array}{l}
\chi^{(3)}(-\omega_1+\omega_2+\omega_3;-\omega_1,\omega_2,\omega_3)+\chi^{(3)}(-\omega_1+\omega_3+\omega_2;-\omega_1,\omega_3,\omega_2) \\
+\chi^{(3)}(\omega_2-\omega_1+\omega_3;\omega_2,\omega_1,\omega_3)+\chi^{(3)}(\omega_2+\omega_3-\omega_1;\omega_2,\omega_3,-\omega_1) \\
+\chi^{(3)}(\omega_3-\omega_1+\omega_2;\omega_3,-\omega_1,\omega_2)+\chi^{(3)}(\omega_3+\omega_2-\omega_1;\omega_3,\omega_2,-\omega_1)
\end{array}\right]\mathrm{E}^*(\omega_1)\mathrm{E}(\omega_2)\mathrm{E}(\omega_3)\left\langle e^{-i\omega_1 t}\middle| e^{-i(-\omega_1+\omega_2+\omega_3)t}\right\rangle \\
+\left[\begin{array}{l}
\chi^{(3)}(\omega_1-\omega_2+\omega_3;\omega_1,-\omega_2,\omega_3)+\chi^{(3)}(\omega_1+\omega_3-\omega_2;\omega_1,\omega_3,-\omega_2) \\
+\chi^{(3)}(-\omega_2+\omega_1+\omega_3;-\omega_2,\omega_1,\omega_3)+\chi^{(3)}(-\omega_2+\omega_3+\omega_1;-\omega_2,\omega_3,\omega_1) \\
+\chi^{(3)}(\omega_3+\omega_1-\omega_2;\omega_3,\omega_1,\omega_2)+\chi^{(3)}(\omega_3-\omega_2+\omega_1;\omega_3,-\omega_2,\omega_1)
\end{array}\right]\mathrm{E}(\omega_1)\mathrm{E}^*(\omega_2)\mathrm{E}(\omega_3)\left\langle e^{-i\omega_2 t}\middle| e^{-i(\omega_1-\omega_2+\omega_3)t}\right\rangle \\
+\left[\begin{array}{l}
\chi^{(3)}(\omega_1+\omega_2-\omega_3;\omega_1,\omega_2,-\omega_3)+\chi^{(3)}(\omega_1-\omega_3+\omega_2;\omega_1,-\omega_3,\omega_2) \\
+\chi^{(3)}(\omega_2+\omega_1-\omega_3;\omega_2,\omega_1,-\omega_3)+\chi^{(3)}(\omega_2-\omega_3+\omega_1;\omega_2,-\omega_3,\omega_1) \\
+\chi^{(3)}(-\omega_3+\omega_1+\omega_2;-\omega_3,\omega_1,\omega_2)+\chi^{(3)}(-\omega_3+\omega_2+\omega_1;-\omega_3,\omega_2,\omega_1)
\end{array}\right]\mathrm{E}(\omega_1)\mathrm{E}(\omega_2)\mathrm{E}^*(\omega_3)\left\langle e^{-i\omega_3 t}\middle| e^{-i(\omega_1+\omega_2-\omega_3)t}\right\rangle
\end{array}\right\}
$$

$$
+\frac{1}{27}\left\{\begin{array}{l}
\left[\chi^{(3)}(\omega_1;\omega_1,\omega_1,-\omega_1)+\chi^{(3)}(\omega_1;\omega_1,-\omega_1,\omega_1)+\chi^{(3)}(\omega_1;-\omega_1,\omega_1,\omega_1)\right]\mathrm{E}(\omega_1)\left|\mathrm{E}(\omega_1)\right|^2 e^{-i\omega_1 t} \\
+\left[\chi^{(3)}(\omega_2;\omega_2,\omega_2,-\omega_2)+\chi^{(3)}(\omega_2;\omega_2,-\omega_2,\omega_2)+\chi^{(3)}(\omega_2;-\omega_2,\omega_2,\omega_2)\right]\mathrm{E}(\omega_2)\left|\mathrm{E}(\omega_2)\right|^2 e^{-i\omega_2 t} \\
+\left[\chi^{(3)}(\omega_3;\omega_3,\omega_3,-\omega_3)+\chi^{(3)}(\omega_3;\omega_3,-\omega_3,\omega_3)+\chi^{(3)}(\omega_3;-\omega_3,\omega_3,\omega_3)\right]\mathrm{E}(\omega_3)\left|\mathrm{E}(\omega_3)\right|^2 e^{-i\omega_3 t}
\end{array}\right\}
$$

$$
+\frac{1}{27}\left\{\begin{array}{l}
+\left[\begin{array}{l}
\chi^{(3)}(\omega_2;-\omega_1,\omega_1,\omega_2)+\chi^{(3)}(\omega_2;\omega_1,-\omega_1,\omega_2)+\chi^{(3)}(\omega_2;\omega_2,\omega_1,-\omega_1) \\
+\chi^{(3)}(\omega_2;\omega_2,-\omega_1,\omega_1)+\chi^{(3)}(\omega_2;\omega_1,\omega_2,-\omega_1)+\chi^{(3)}(\omega_2;-\omega_1,\omega_2,\omega_1)
\end{array}\right]\mathrm{E}(\omega_2)\left|\mathrm{E}(\omega_1)\right|^2 e^{-i\omega_2 t} \\
+\left[\begin{array}{l}
\chi^{(3)}(\omega_3;-\omega_1,\omega_1,\omega_3)+\chi^{(3)}(\omega_3;\omega_1,-\omega_1,\omega_3)+\chi^{(3)}(\omega_3;\omega_3,\omega_1,-\omega_1) \\
+\chi^{(3)}(\omega_3;\omega_3,-\omega_1,\omega_1)+\chi^{(3)}(\omega_3;\omega_1,\omega_3,-\omega_1)+\chi^{(3)}(\omega_3;-\omega_1,\omega_3,\omega_1)
\end{array}\right]\mathrm{E}(\omega_1)\left|\mathrm{E}(\omega_1)\right|^2 e^{-i\omega_3 t} \\
+\left[\begin{array}{l}
\chi^{(3)}(\omega_1;-\omega_2,\omega_2,\omega_1)+\chi^{(3)}(\omega_1;\omega_2,-\omega_2,\omega_1)+\chi^{(3)}(\omega_1;\omega_1,\omega_2,-\omega_2) \\
+\chi^{(3)}(\omega_1;\omega_1,-\omega_2,\omega_2)+\chi^{(3)}(\omega_1;\omega_2,\omega_1,-\omega_2)+\chi^{(3)}(\omega_1;-\omega_2,\omega_1,\omega_2)
\end{array}\right]\mathrm{E}(\omega_1)\left|\mathrm{E}(\omega_2)\right|^2 e^{-i\omega_1 t} \\
+\left[\begin{array}{l}
\chi^{(3)}(\omega_3;-\omega_2,\omega_2,\omega_3)+\chi^{(3)}(\omega_3;\omega_2,-\omega_2,\omega_3)+\chi^{(3)}(\omega_3;\omega_3,\omega_2,-\omega_2) \\
+\chi^{(3)}(\omega_3;\omega_3,-\omega_2,\omega_2)+\chi^{(3)}(\omega_3;\omega_2,\omega_3,-\omega_2)+\chi^{(3)}(\omega_3;-\omega_2,\omega_3,\omega_2)
\end{array}\right]\mathrm{E}(\omega_3)\left|\mathrm{E}(\omega_2)\right|^2 e^{-i\omega_3 t} \\
+\left[\begin{array}{l}
\chi^{(3)}(\omega_1;-\omega_3,\omega_3,\omega_1)+\chi^{(3)}(\omega_1;\omega_3,-\omega_3,\omega_1)+\chi^{(3)}(\omega_1;\omega_1,\omega_3,-\omega_3) \\
+\chi^{(3)}(\omega_1;\omega_1,-\omega_3,\omega_3)+\chi^{(3)}(\omega_1;\omega_3,\omega_1,-\omega_3)+\chi^{(3)}(\omega_1;-\omega_3,\omega_1,\omega_3)
\end{array}\right]\mathrm{E}(\omega_1)\left|\mathrm{E}(\omega_3)\right|^2 e^{-i\omega_1 t} \\
+\left[\begin{array}{l}
\chi^{(3)}(\omega_2;-\omega_3,\omega_3,\omega_2)+\chi^{(3)}(\omega_2;\omega_3,-\omega_3,\omega_2)+\chi^{(3)}(\omega_2;\omega_2,\omega_3,-\omega_3) \\
+\chi^{(3)}(\omega_2;\omega_2,-\omega_3,\omega_3)+\chi^{(3)}(\omega_2;\omega_3,\omega_2,-\omega_3)+\chi^{(3)}(\omega_2;-\omega_3,\omega_2,\omega_3)
\end{array}\right]\mathrm{E}(\omega_2)\left|\mathrm{E}(\omega_3)\right|^2 e^{-i\omega_2 t}
\end{array}\right\}
$$

$$
+c.c. \tag{3.2}
$$

Present address: Photonic Sciences Lab, Physical Research Laboratory, Navarangpura, Ahmedabad 380009, Gujarat, India, email: asvrao@prl.res.in

The combination of three fundamental frequencies produces 210 nonlinear susceptibility tensors. In the Frequency Generation (FG) terms (first five terms in the Eq. 3.2) are $3\omega_1$, $3\omega_2$, $3\omega_3$, $2\omega_1+\omega_2$, $2\omega_1+\omega_3$,$2\omega_2+\omega_1$, $2\omega_2+\omega_3$, $2\omega_3+\omega_1$, $2\omega_3+\omega_2$, $\omega_1+\omega_2+\omega_3$, $2\omega_1-\omega_2$, $2\omega_1-\omega_3$, $2\omega_2-\omega_1$, $2\omega_2-\omega_3$, $2\omega_3-\omega_1$, $2\omega_3-\omega_2$, $-\omega_1+\omega_2+\omega_3$, $\omega_1-\omega_2+\omega_3$, and $\omega_1+\omega_2-\omega_3$. The addition of three frequencies rise to give third frequency is THG and number of elements in the susceptibility tensor of THG is 486 {[(81+81(c.c.)) ($3\omega_1$)]+[(81+81(c.c.)) ($3\omega_2$)]+[(81+81(c.c.)) ($3\omega_3$)]}. Anyone of the frequencies can generate SHG and then it will sum with another frequency to create SFSHG. Here number of elements in the susceptibility tensor is 2,916 {[3(81+81(c.c.)) ($2\omega_1+\omega_2$)]+[3(81+81(c.c.)) ($2\omega_1+\omega_3$)]+[3(81+81(c.c.)) ($2\omega_2+\omega_1$)]+[3(81+81(c.c.)) ($2\omega_2+\omega_3$)]+[3(81+81(c.c.)) ($2\omega_3+\omega_1$)]+[3(81+81(c.c.)) ($2\omega_3+\omega_2$)]}. Three frequencies can add to create the SFG. Number of elements in the susceptibility tensor of SFG is 972 {6(81+81(c.c.)) ($\omega_1+\omega_2+\omega_3$)}.SHG of one frequency can subtract with another frequency to produce DFSHG and the number of elements in the susceptibility tensor is 2,916 {[3(81+81(c.c.)) ($2\omega_1-\omega_2$)]+[3(81+81(c.c.)) ($2\omega_1-\omega_3$)]+[3(81+81(c.c.)) ($2\omega_2-\omega_1$)]+[3(81+81(c.c.)) ($2\omega_2-\omega_3$)]+[3(81+81(c.c.)) ($2\omega_3-\omega_1$)]+[3(81+81(c.c.)) ($2\omega_3-\omega_2$)]}.The three frequencies can combine as sum with difference to produce DFSFG and the number of elements in the susceptibility tensor is 2,916 {[6(81+81(c.c.))($-\omega_1+\omega_2+\omega_3$)]+[6(81+81(c.c.)) ($\omega_1-\omega_2+\omega_3$)]+[6(81+81(c.c.))($\omega_1+\omega_2-\omega_3$)]}.Another one is No Frequency Generation (NFG), whose output frequencies are same as fundamental frequencies ($\omega_1$, $\omega_2$, and $\omega_3$)and these terms can be seen in the last set of Eq. 3.2. NFG terms themselves shows different nonlinear behavior depending upon the interacting material. At the end of the next section, we have given a detailed explanation for this scenario. For this section, we have used NFG as suffix for D-matrix representation of these terms by two ways: DNFG (Degenerative No Frequency Generation) for the case of generating the optical nonlinearity by single fundamental frequency and NDNFG (Non Degenerative No Frequency Generation) where more than one fundamental frequency will involve in the nonlinear process. The number of elements in the tensor matrix of DNFG is 1,458 {[3(81+81(c.c.)) ($\omega_1$)]+[3(81+81(c.c.)) ($\omega_2$)]+[3(81+81(c.c.)) ($\omega_3$)]}.In the case of NDNFG, number of elements in the tensor matrix is 5,346 {[12(81+81(c.c.)) ($\omega_1$)]+[12(81+81(c.c.)) ($\omega_2$)]+[12(81+81(c.c.)) ($\omega_3$)]}.Therefore the total number of elements is 17,270.

### *3.1.Third Harmonic Generation*

Let us consider beam with single frequency ω propagating in the material which has centro-symmetric nature then its third-order nonlinear polarization contains the THG, SFSHG, DFSHG, DFSFG, and NFG. In this process, SFSHG term illicitly becomes THG due to degenerative nature of the fundamental frequency and thus asset in increasing the probability of THG. The next DFSHG and DFSFG will give the fundamental frequency as a consequence of degeneracy of fundamental frequency and their contribution in the energy transfer from ω to 3ω will be zero. But due to their internal conversion, the time in producing the THG may consume and time of THG is restricted to given interacting optical length of material. Under phase matching condition $\hat{k}_{3\omega}=3\hat{k}_{\omega}$, THG can be enriched. This matching restricts the energy flow in the ω→3ω direction. The incident optical field amplitude (equation 2.3) for THG gives the nonlinear polarization as

$$\begin{aligned}P^{(3)}(t) &= \chi^{(3)}(3\omega;\omega,\omega,\omega)E^3(\omega)e^{-i3\omega t}\\ &+\left[\chi^{(3)}(\omega;\omega,\omega,-\omega)+\chi^{(3)}(\omega;\omega,-\omega,\omega)+\chi^{(3)}(\omega;-\omega,\omega,\omega)\right]\left|E(\omega)\right|^2 E(\omega)e^{-i\omega t}+\text{c.c.}\end{aligned} \tag{3.3}$$

Present address: Photonic Sciences Lab, Physical Research Laboratory, Navarangpura, Ahmedabad 380009, Gujarat, India, email: asvrao@prl.res.in

The number of susceptibility elements in the THG is 648 {[81+81(c.c.) (3ω)]+3(81+81(c.c.)) (ω)]}. In off resonant process the susceptibility tensor can be written in the Kleinman's D-matrix form. The general expression for THG is

$$P^{(3)}(t) = 3d_{3\omega}^{\mathrm{THG}} E^3(\omega)e^{-i3\omega t} + 9d_{\omega}^{\mathrm{DNFG}} |E(\omega)|^2 E(\omega)e^{-i\omega t} \tag{3.4}$$

$$\begin{bmatrix} P_1(3\omega) \\ P_2(3\omega) \\ P_3(3\omega) \end{bmatrix} = \frac{1}{81}\begin{pmatrix} d_{11} & d_{12} & d_{13} & d_{14} & d_{15} & d_{16} & d_{17} & d_{18} & d_{19} & d_{1,10} \\ d_{21} & d_{22} & d_{23} & d_{24} & d_{25} & d_{26} & d_{27} & d_{28} & d_{29} & d_{2,10} \\ d_{31} & d_{32} & d_{33} & d_{34} & d_{35} & d_{36} & d_{37} & d_{38} & d_{39} & d_{3,10} \end{pmatrix}\begin{pmatrix} E_1^3(\omega) \\ E_2^3(\omega) \\ E_3^3(\omega) \\ 3E_2(\omega)E_3(\omega)E_3(\omega) \\ 3E_3(\omega)E_2(\omega)E_2(\omega) \\ 3E_1(\omega)E_3(\omega)E_3(\omega) \\ 3E_1(\omega)E_1(\omega)E_3(\omega) \\ 3E_1(\omega)E_2(\omega)E_2(\omega) \\ 3E_1(\omega)E_1(\omega)E_2(\omega) \\ 6E_1(\omega)E_2(\omega)E_3(\omega) \end{pmatrix} \tag{3.5}$$

$$\begin{bmatrix} P_1(\omega) \\ P_2(\omega) \\ P_3(\omega) \end{bmatrix} = \frac{1}{81}\begin{pmatrix} d_{11} & d_{12} & d_{13} & d_{14} & d_{15} & d_{16} & d_{17} & d_{18} & d_{19} & d_{1,10} \\ d_{21} & d_{22} & d_{23} & d_{24} & d_{25} & d_{26} & d_{27} & d_{28} & d_{29} & d_{2,10} \\ d_{31} & d_{32} & d_{33} & d_{34} & d_{35} & d_{36} & d_{37} & d_{38} & d_{39} & d_{3,10} \end{pmatrix}\begin{pmatrix} |E_1(\omega)|^2 E_1(\omega) \\ |E_2(\omega)|^2 E_2(\omega) \\ |E_3(\omega)|^2 E_3(\omega) \\ 3|E_3(\omega)|^2 E_2(\omega) \\ 3|E_2(\omega)|^2 E_3(\omega) \\ 3|E_3(\omega)|^2 E_1(\omega) \\ 3|E_1(\omega)|^2 E_3(\omega) \\ 3|E_2(\omega)|^2 E_1(\omega) \\ 3|E_1(\omega)|^2 E_2(\omega) \\ 6E_1(\omega)E_2(\omega)E_3^*(\omega) \end{pmatrix} \tag{3.6}$$

Total number of independent terms is 60 {[30 (3ω)]+[30 (ω)]}. In far off-resonant case the D-matrix becomes

$$d = \frac{1}{81}\begin{pmatrix} d_{11} & d_{12} & d_{13} & d_{14} & d_{15} & d_{16} & d_{17} & d_{18} & d_{19} & d_{1,10} \\ d_{19} & d_{22} & d_{23} & d_{24} & d_{25} & d_{14} & d_{1,10} & d_{12} & d_{18} & d_{15} \\ d_{17} & d_{25} & d_{33} & d_{23} & d_{24} & d_{13} & d_{16} & d_{15} & d_{1,10} & d_{14} \end{pmatrix} \tag{3.7}$$

Number of independent variables is 30 {[15 (3ω)]+[15 (ω)]}.

For producing SFSHG, DFSHG, we require two optical frequencies in the light-matter interaction. Let two beams of frequencies $\omega_1$ and $\omega_2$with equal intensities are interacting in the nonlinear media and the optical field amplitude is given by equation 2.1. The general form of third-order susceptibility is

$$P^{(3)}(t) = \frac{1}{8}\left\{\chi^{(3)}(3\omega_1;\omega_1,\omega_1,\omega_1)E^3(\omega_1)e^{-i3\omega_1 t} + \chi^{(3)}(3\omega_2;\omega_2,\omega_2,\omega_2)E^3(\omega_2)e^{-i3\omega_2 t}\right\}$$

$$+\frac{1}{8}\left\{\begin{matrix}\left[\chi^{(3)}(2\omega_1+\omega_2;\omega_1,\omega_1,\omega_2) + \chi^{(3)}(\omega_1+\omega_2+\omega_1;\omega_1,\omega_2,\omega_1) + \chi^{(3)}(\omega_2+2\omega_1;\omega_2,\omega_1,\omega_1)\right]E^2(\omega_1)E(\omega_2)e^{-i(2\omega_1+\omega_2)t} \\ +\left[\chi^{(3)}(\omega_1+2\omega_2;\omega_1,\omega_2,\omega_2) + \chi^{(3)}(\omega_2+\omega_1+\omega_2;\omega_2,\omega_1,\omega_2) + \chi^{(3)}(2\omega_2+\omega_1;\omega_2,\omega_2,\omega_1)\right]E(\omega_1)E^2(\omega_2)e^{-i(\omega_1+2\omega_2)t}\end{matrix}\right\}$$

$$+\frac{1}{8}\left\{\begin{matrix}\left[\chi^{(3)}(2\omega_1-\omega_2;\omega_1,\omega_1,-\omega_2) + \chi^{(3)}(\omega_1-\omega_2+\omega_1;\omega_1,-\omega_2,\omega_1) + \chi^{(3)}(-\omega_2+2\omega_1;-\omega_2,\omega_1,\omega_1)\right]E^2(\omega_1)E^*(\omega_2)\left\langle e^{-i\omega_2 t}\middle| e^{-i(2\omega_1-\omega_2)t}\right\rangle \\ +\left[\chi^{(3)}(-\omega_1+2\omega_2;-\omega_1,\omega_2,\omega_2) + \chi^{(3)}(\omega_2-\omega_1+\omega_2;\omega_2,-\omega_1,\omega_2) + \chi^{(3)}(2\omega_2-\omega_1;\omega_2,\omega_2,-\omega_1)\right]E^*(\omega_1)E^2(\omega_2)\left\langle e^{-i\omega_1 t}\middle| e^{-i(-\omega_1+2\omega_2)t}\right\rangle\end{matrix}\right\}$$

Present address: Photonic Sciences Lab, Physical Research Laboratory, Navarangpura, Ahmedabad 380009, Gujarat, India, email: asvrao@prl.res.in

$$
+\frac{1}{8}\left\{\begin{array}{l}
\left[\chi^{(3)}(\omega_1;\omega_1,\omega_1,-\omega_1)+\chi^{(3)}(\omega_1;\omega_1,-\omega_1,\omega_1)+\chi^{(3)}(\omega_1;-\omega_1,\omega_1,\omega_1)\right]\mathrm{E}(\omega_1)|\mathrm{E}(\omega_1)|^2 e^{-i\omega_1 t}\\
+\left[\chi^{(3)}(\omega_2;\omega_2,\omega_2,-\omega_2)+\chi^{(3)}(\omega_2;\omega_2,-\omega_2,\omega_2)+\chi^{(3)}(\omega_2;-\omega_2,\omega_2,\omega_2)\right]\mathrm{E}(\omega_2)|\mathrm{E}(\omega_2)|^2 e^{-i\omega_2 t}\\
+\left[\begin{array}{l}\chi^{(3)}(\omega_2;-\omega_1,\omega_1,\omega_2)+\chi^{(3)}(\omega_2;\omega_1,-\omega_1,\omega_2)+\chi^{(3)}(\omega_2;\omega_2,\omega_1,-\omega_1)\\+\chi^{(3)}(\omega_2;\omega_2,-\omega_1,\omega_1)+\chi^{(3)}(\omega_2;\omega_1,\omega_2,-\omega_1)+\chi^{(3)}(\omega_2;-\omega_1,\omega_2,\omega_1)\end{array}\right]\mathrm{E}(\omega_2)|\mathrm{E}(\omega_1)|^2 e^{-i\omega_2 t}\\
+\left[\begin{array}{l}\chi^{(3)}(\omega_1;-\omega_2,\omega_2,\omega_1)+\chi^{(3)}(\omega_1;\omega_2,-\omega_2,\omega_1)+\chi^{(3)}(\omega_1;\omega_1,\omega_2,-\omega_2)\\+\chi^{(3)}(\omega_1;\omega_1,-\omega_2,\omega_2)+\chi^{(3)}(\omega_1;\omega_2,\omega_1,-\omega_2)+\chi^{(3)}(\omega_1;-\omega_2,\omega_1,\omega_2)\end{array}\right]\mathrm{E}(\omega_1)|\mathrm{E}(\omega_2)|^2 e^{-i\omega_1 t}
\end{array}\right\}
$$

$$
+c.c. \tag{3.8}
$$

The combination of two fundamental frequencies produces 64 nonlinear terms in the following nonlinear processes. First one is THG of $\omega_1$and$\omega_2$, where the number of elements in the susceptibility tensors is 324 {[(81+81(c.c.)) ($3\omega_1$)]+[(81+81(c.c.)) ($3\omega_2$)]}. Next one is SFSHG, here number of elements in the susceptibility tensors is 972 {[3(81+81(c.c.)) ($2\omega_1+\omega_2$)]+[3(81+81(c.c.)) ($2\omega_2+\omega_1$)]}. The third frequency generation term is DFSHG and number of elements in the susceptibility tensors is 972 {[3(81+81(c.c.)) ($2\omega_1-\omega_2$)]+[3(81+81(c.c.)) ($2\omega_2-\omega_1$)]}. In DNFG process, number of elements in the susceptibility tensors is 972 {[3(81+81(c.c.)) ($\omega_1$)]+[3(81+81(c.c.)) $\omega_2$)]}.In NDNFG process, number of elements in the susceptibility tensors is 1,944 {[6(81+81(c.c.)) ($\omega_1$)]+[6(81+81(c.c.)) $\omega_2$)]}. Therefore total number of elements is 5,184.

*3.2.SFSHG*

To generate SFSHG, crystal has to satisfy the phase matching condition$\hat{k}_{2\omega 1+\omega 2}=2\hat{k}_{2\omega 1}+\hat{k}_{\omega 2}$. This matching restricts the energy transfer in the ($\omega_1$, $\omega_2$) →2 $\omega_1+\omega_2$ direction. The nonlinear polarization is

$$
\mathrm{P}^{(3)}(t)=+\frac{1}{8}\left\{\left[\begin{array}{l}\chi^{(3)}(2\omega_1+\omega_2;\omega_1,\omega_1,\omega_2)+\chi^{(3)}(\omega_1+\omega_2+\omega_1;\omega_1,\omega_2,\omega_1)\\+\chi^{(3)}(\omega_2+2\omega_1;\omega_2,\omega_1,\omega_1)\end{array}\right]\mathrm{E}^2(\omega_1)\mathrm{E}(\omega_2)e^{-i(\omega_2+2\omega_1)t}\right\}
$$

$$
+\frac{1}{8}\left\{\begin{array}{l}
\left[\chi^{(3)}(\omega_1;\omega_1,\omega_1,-\omega_1)+\chi^{(3)}(\omega_1;\omega_1,-\omega_1,\omega_1)+\chi^{(3)}(\omega_1;-\omega_1,\omega_1,\omega_1)\right]\mathrm{E}(\omega_1)|\mathrm{E}(\omega_1)|^2 e^{-i\omega_1 t}\\
+\left[\chi^{(3)}(\omega_2;\omega_2,\omega_2,-\omega_2)+\chi^{(3)}(\omega_2;\omega_2,-\omega_2,\omega_2)+\chi^{(3)}(\omega_2;-\omega_2,\omega_2,\omega_2)\right]\mathrm{E}(\omega_2)|\mathrm{E}(\omega_2)|^2 e^{-i\omega_2 t}\\
+\left[\begin{array}{l}\chi^{(3)}(\omega_2;-\omega_1,\omega_1,\omega_2)+\chi^{(3)}(\omega_2;\omega_1,-\omega_1,\omega_2)+\chi^{(3)}(\omega_2;\omega_2,\omega_1,-\omega_1)\\+\chi^{(3)}(\omega_2;\omega_2,-\omega_1,\omega_1)+\chi^{(3)}(\omega_2;\omega_1,\omega_2,-\omega_1)+\chi^{(3)}(\omega_2;-\omega_1,\omega_2,\omega_1)\end{array}\right]\mathrm{E}(\omega_2)|\mathrm{E}(\omega_1)|^2 e^{-i\omega_2 t}\\
+\left[\begin{array}{l}\chi^{(3)}(\omega_1;-\omega_2,\omega_2,\omega_1)+\chi^{(3)}(\omega_1;\omega_2,-\omega_2,\omega_1)+\chi^{(3)}(\omega_1;\omega_1,\omega_2,-\omega_2)\\+\chi^{(3)}(\omega_1;\omega_1,-\omega_2,\omega_2)+\chi^{(3)}(\omega_1;\omega_2,\omega_1,-\omega_2)+\chi^{(3)}(\omega_1;-\omega_2,\omega_1,\omega_2)\end{array}\right]\mathrm{E}(\omega_1)|\mathrm{E}(\omega_2)|^2 e^{-i\omega_1 t}
\end{array}\right\}
$$

$$
+c.c. \tag{3.9}
$$

Here total number of susceptibility tensors is 64 {32+32(c.c.) =64}. Susceptibility tensor of SFSHG contains486 {[3(81+81(c.c.)) ($2\omega_1+\omega_2$)]}number of elements. In DNFG process, number of elements in the susceptibility tensors is 972 {[3(81+81(c.c.)) ($\omega_1$)]+[3(81+81(c.c.)) $\omega_2$)]} and In NDNFG is 1,944 {[6(81+81(c.c.)) ($\omega_1$)]+[6(81+81(c.c.)) $\omega_2$)]}. Therefore total number of elements is 3,402. In off resonant, we can write the nonlinear polarization as

$$
\mathrm{P}^{(3)}(t)=\frac{9}{8}d^{SFSHG}_{2\omega_1+\omega_2}\mathrm{E}^2(\omega_1)\mathrm{E}(\omega_2)e^{-i(2\omega_1+\omega_2)t}+\frac{9}{8}\left\{d^{DNFG}_{\omega_1}\mathrm{E}(\omega_1)|\mathrm{E}(\omega_1)|^2 e^{-i\omega_1 t}+d^{DNFG}_{\omega_2}\mathrm{E}(\omega_2)|\mathrm{E}(\omega_2)|^2 e^{-i\omega_2 t}\right\}
$$
$$
+\frac{9}{4}\left\{d^{NDNFG}_{\omega_1}\mathrm{E}(\omega_1)|\mathrm{E}(\omega_2)|^2 e^{-i\omega_1 t}+d^{NDNFG}_{\omega_2}\mathrm{E}(\omega_2)|\mathrm{E}(\omega_1)|^2 e^{-i\omega_2 t}\right\} \tag{3.10}
$$

Here number of D-matrix tensors is 5.

Present address: Photonic Sciences Lab, Physical Research Laboratory, Navarangpura, Ahmedabad 380009, Gujarat, India, email: asvrao@prl.res.in

$$\begin{bmatrix} P_1(2\omega_1+\omega_2) \\ P_2(2\omega_1+\omega_2) \\ P_3(2\omega_1+\omega_2) \end{bmatrix} = \frac{1}{81}\begin{pmatrix} d_{11} & d_{12} & d_{13} & d_{14} & d_{15} & d_{16} & d_{17} & d_{18} & d_{19} & d_{1,10} \\ d_{21} & d_{22} & d_{23} & d_{25} & d_{26} & d_{26} & d_{27} & d_{28} & d_{29} & d_{2,10} \\ d_{31} & d_{32} & d_{33} & d_{34} & d_{35} & d_{36} & d_{37} & d_{38} & d_{39} & d_{3,10} \end{pmatrix} \begin{pmatrix} E_1^2(\omega_1)E_1(\omega_2) \\ E_2^2(\omega_1)E_2(\omega_2) \\ E_2^2(\omega_1)E_2(\omega_2) \\ 3E_2(\omega_1)E_3(\omega_1)E_3(\omega_2) \\ 3E_3(\omega_1)E_2(\omega_1)E_2(\omega_2) \\ 3E_1(\omega_1)E_3(\omega_1)E_3(\omega_2) \\ 3E_1(\omega_1)E_1(\omega_1)E_3(\omega_2) \\ 3E_2(\omega_1)E_2(\omega_1)E_2(\omega_2) \\ 3E_1(\omega_1)E_1(\omega_1)E_2(\omega_2) \\ 6E_1(\omega_1)E_2(\omega_1)E_3(\omega_2) \end{pmatrix} \tag{3.11}$$

$$\begin{bmatrix} P_1(\omega_1) \\ P_2(\omega_1) \\ P_3(\omega_1) \end{bmatrix} = \frac{1}{81}\begin{pmatrix} d_{11} & d_{12} & d_{13} & d_{14} & d_{15} & d_{16} & d_{17} & d_{18} & d_{19} & d_{1,10} \\ d_{21} & d_{22} & d_{23} & d_{25} & d_{26} & d_{26} & d_{27} & d_{28} & d_{29} & d_{2,10} \\ d_{31} & d_{32} & d_{33} & d_{34} & d_{35} & d_{36} & d_{37} & d_{38} & d_{39} & d_{3,10} \end{pmatrix} \begin{pmatrix} |E_1(\omega_1)|^2 E_1(\omega_1) \\ |E_2(\omega_1)|^2 E_2(\omega_1) \\ |E_3(\omega_1)|^2 E_3(\omega_1) \\ 3|E_3(\omega_1)|^2 E_2(\omega_1) \\ 3|E_2(\omega_1)|^2 E_3(\omega_1) \\ 3|E_3(\omega_1)|^2 E_1(\omega_1) \\ 3|E_1(\omega_1)|^2 E_3(\omega_1) \\ 3|E_2(\omega_1)|^2 E_1(\omega_1) \\ 3|E_1(\omega_1)|^2 E_2(\omega_1) \\ 6E_1(\omega_1)E_2(\omega_1)E_3^*(\omega_1) \end{pmatrix} \tag{3.12}$$

$$\begin{bmatrix} P_1(\omega_2) \\ P_2(\omega_2) \\ P_3(\omega_2) \end{bmatrix} = \frac{1}{81}\begin{pmatrix} d_{11} & d_{12} & d_{13} & d_{14} & d_{15} & d_{16} & d_{17} & d_{18} & d_{19} & d_{1,10} \\ d_{21} & d_{22} & d_{23} & d_{25} & d_{26} & d_{26} & d_{27} & d_{28} & d_{29} & d_{2,10} \\ d_{31} & d_{32} & d_{33} & d_{34} & d_{35} & d_{36} & d_{37} & d_{38} & d_{39} & d_{3,10} \end{pmatrix} \begin{pmatrix} |E_1(\omega_2)|^2 E_1(\omega_2) \\ |E_2(\omega_2)|^2 E_2(\omega_2) \\ |E_3(\omega_2)|^2 E_3(\omega_2) \\ 3|E_3(\omega_2)|^2 E_2(\omega_2) \\ 3|E_2(\omega_2)|^2 E_3(\omega_2) \\ 3|E_3(\omega_2)|^2 E_1(\omega_2) \\ 3|E_1(\omega_2)|^2 E_3(\omega_2) \\ 3|E_2(\omega_2)|^2 E_1(\omega_2) \\ 3|E_1(\omega_2)|^2 E_2(\omega_2) \\ 6E_1(\omega_2)E_2(\omega_2)E_3^*(\omega_2) \end{pmatrix} \tag{3.13}$$

$$\begin{bmatrix} P_1(\omega_1) \\ P_2(\omega_1) \\ P_3(\omega_1) \end{bmatrix} = \frac{1}{81}\begin{pmatrix} d_{11} & d_{12} & d_{13} & d_{14} & d_{15} & d_{16} & d_{17} & d_{18} & d_{19} & d_{1,10} \\ d_{21} & d_{22} & d_{23} & d_{25} & d_{26} & d_{26} & d_{27} & d_{28} & d_{29} & d_{2,10} \\ d_{31} & d_{32} & d_{33} & d_{34} & d_{35} & d_{36} & d_{37} & d_{38} & d_{39} & d_{3,10} \end{pmatrix} \begin{pmatrix} |E_1(\omega_2)|^2 E_1(\omega_1) \\ |E_2(\omega_2)|^2 E_2(\omega_1) \\ |E_3(\omega_2)|^2 E_3(\omega_1) \\ 3|E_3(\omega_2)|^2 E_2(\omega_1) \\ 3|E_2(\omega_2)|^2 E_3(\omega_1) \\ 3|E_3(\omega_2)|^2 E_1(\omega_1) \\ 3|E_1(\omega_2)|^2 E_3(\omega_1) \\ 3|E_2(\omega_2)|^2 E_1(\omega_1) \\ 3|E_1(\omega_2)|^2 E_2(\omega_1) \\ 6E_1(\omega_2)E_2(\omega_1)E_3^*(\omega_2) \end{pmatrix} \tag{3.14}$$

Present address: Photonic Sciences Lab, Physical Research Laboratory, Navarangpura, Ahmedabad 380009, Gujarat, India, email: asvrao@prl.res.in

$$
\begin{bmatrix} P_1(\omega_2) \\ P_2(\omega_2) \\ P_3(\omega_2) \end{bmatrix} = \frac{1}{81} \begin{pmatrix} d_{11} & d_{12} & d_{13} & d_{14} & d_{15} & d_{16} & d_{17} & d_{18} & d_{19} & d_{1,10} \\ d_{21} & d_{22} & d_{23} & d_{25} & d_{26} & d_{26} & d_{27} & d_{28} & d_{29} & d_{2,10} \\ d_{31} & d_{32} & d_{33} & d_{34} & d_{35} & d_{36} & d_{37} & d_{38} & d_{39} & d_{3,10} \end{pmatrix} \begin{pmatrix} |E_1(\omega_1)|^2 E_1(\omega_2) \\ |E_2(\omega_1)|^2 E_2(\omega_2) \\ |E_3(\omega_1)|^2 E_3(\omega_2) \\ 3|E_3(\omega_1)|^2 E_2(\omega_2) \\ 3|E_2(\omega_1)|^2 E_3(\omega_2) \\ 3|E_3(\omega_1)|^2 E_1(\omega_2) \\ 3|E_1(\omega_1)|^2 E_3(\omega_2) \\ 3|E_2(\omega_1)|^2 E_1(\omega_2) \\ 3|E_1(\omega_1)|^2 E_2(\omega_2) \\ 6E_1(\omega_1)E_2(\omega_2)E_3^*(\omega_1) \end{pmatrix} \tag{3.15}
$$

Total number of D-matrix elements is 150:{[30 (SFSHG)]+[60 (DNFG)]+[60 (NDNFG)]}. In case of far off-resonance number of independent D-matrix elements becomes 75. Here we have not discussed DFSHG because its polarization expression follows the same as SFSHG. From equations 3.4 and 3.10, the probability of producing THG is 8/3 times larger than the SFSHG and DFSHG.

*3.3.SFDFG*

To produce SFDFG, we must make three frequencies to interact within the medium. Let us fix the phase matching condition $\hat{k}_{\omega1+\omega2-\omega3}=\hat{k}_{\omega1}+\hat{k}_{\omega2}-\hat{k}_{\omega3}$ (assumption: $\omega_1+\omega_2>\omega_3$) then from equation 3.2 the expression for nonlinear polarization is

$$
\begin{aligned}
P^{(3)}(t) = \frac{1}{27} & \begin{bmatrix} \chi^{(3)}(\omega_1+\omega_2-\omega_3;\omega_1,\omega_2,-\omega_3) + \chi^{(3)}(\omega_1-\omega_3+\omega_2;\omega_1,-\omega_3,\omega_2) \\ +\chi^{(3)}(\omega_2+\omega_1-\omega_3;\omega_2,\omega_1,-\omega_3) + \chi^{(3)}(\omega_2-\omega_3+\omega_1;\omega_2,-\omega_3,\omega_1) \\ +\chi^{(3)}(-\omega_3+\omega_1+\omega_2;-\omega_3,\omega_1,\omega_2) + \chi^{(3)}(-\omega_3+\omega_2+\omega_1;-\omega_3,\omega_2,\omega_1) \end{bmatrix} E(\omega_1)E(\omega_2)E^*(\omega_3)\left\langle e^{-i\omega_3 t} \middle| e^{-i(\omega_1+\omega_2-\omega_3)t} \right\rangle \\
+\frac{1}{27} & \begin{Bmatrix} \left[\chi^{(3)}(\omega_1;\omega_1,\omega_1,-\omega_1) + \chi^{(3)}(\omega_1;\omega_1,-\omega_1,\omega_1) + \chi^{(3)}(\omega_1;-\omega_1,\omega_1,\omega_1)\right] E(\omega_1)|E(\omega_1)|^2 e^{-i\omega_1 t} \\ +\left[\chi^{(3)}(\omega_2;\omega_2,\omega_2,-\omega_2) + \chi^{(3)}(\omega_2;\omega_2,-\omega_2,\omega_2) + \chi^{(3)}(\omega_2;-\omega_2,\omega_2,\omega_2)\right] E(\omega_2)|E(\omega_2)|^2 e^{-i\omega_2 t} \\ +\left[\chi^{(3)}(\omega_3;\omega_3,\omega_3,-\omega_3) + \chi^{(3)}(\omega_3;\omega_3,-\omega_3,\omega_3) + \chi^{(3)}(\omega_3;-\omega_3,\omega_3,\omega_3)\right] E(\omega_3)|E(\omega_3)|^2 e^{-i\omega_3 t} \end{Bmatrix} \\
+\frac{1}{27} & \begin{Bmatrix} +\begin{bmatrix} \chi^{(3)}(\omega_2;-\omega_1,\omega_1,\omega_2) + \chi^{(3)}(\omega_2;\omega_1,-\omega_1,\omega_2) + \chi^{(3)}(\omega_2;\omega_2,\omega_1,-\omega_1) \\ +\chi^{(3)}(\omega_2;\omega_2,-\omega_1,\omega_1) + \chi^{(3)}(\omega_2;\omega_1,\omega_2,-\omega_1) + \chi^{(3)}(\omega_2;-\omega_1,\omega_2,\omega_1) \end{bmatrix} E(\omega_2)|E(\omega_1)|^2 e^{-i\omega_2 t} \\ +\begin{bmatrix} \chi^{(3)}(\omega_3;-\omega_1,\omega_1,\omega_3) + \chi^{(3)}(\omega_3;\omega_1,-\omega_1,\omega_3) + \chi^{(3)}(\omega_3;\omega_3,\omega_1,-\omega_1) \\ +\chi^{(3)}(\omega_3;\omega_3,-\omega_1,\omega_1) + \chi^{(3)}(\omega_3;\omega_1,\omega_3,-\omega_1) + \chi^{(3)}(\omega_3;-\omega_1,\omega_3,\omega_1) \end{bmatrix} E(\omega_1)|E(\omega_1)|^2 e^{-i\omega_3 t} \\ +\begin{bmatrix} \chi^{(3)}(\omega_1;-\omega_2,\omega_2,\omega_1) + \chi^{(3)}(\omega_1;\omega_2,-\omega_2,\omega_1) + \chi^{(3)}(\omega_1;\omega_1,\omega_2,-\omega_2) \\ +\chi^{(3)}(\omega_1;\omega_1,-\omega_2,\omega_2) + \chi^{(3)}(\omega_1;\omega_2,\omega_1,-\omega_2) + \chi^{(3)}(\omega_1;-\omega_2,\omega_1,\omega_2) \end{bmatrix} E(\omega_1)|E(\omega_2)|^2 e^{-i\omega_1 t} \\ +\begin{bmatrix} \chi^{(3)}(\omega_3;-\omega_2,\omega_2,\omega_3) + \chi^{(3)}(\omega_3;\omega_2,-\omega_2,\omega_3) + \chi^{(3)}(\omega_3;\omega_3,\omega_2,-\omega_2) \\ +\chi^{(3)}(\omega_3;\omega_3,-\omega_2,\omega_2) + \chi^{(3)}(\omega_3;\omega_2,\omega_3,-\omega_2) + \chi^{(3)}(\omega_3;-\omega_2,\omega_3,\omega_2) \end{bmatrix} E(\omega_3)|E(\omega_2)|^2 e^{-i\omega_3 t} \\ +\begin{bmatrix} \chi^{(3)}(\omega_1;-\omega_3,\omega_3,\omega_1) + \chi^{(3)}(\omega_1;\omega_3,-\omega_3,\omega_1) + \chi^{(3)}(\omega_1;\omega_1,\omega_3,-\omega_3) \\ +\chi^{(3)}(\omega_1;\omega_1,-\omega_3,\omega_3) + \chi^{(3)}(\omega_1;\omega_3,\omega_1,-\omega_3) + \chi^{(3)}(\omega_1;-\omega_3,\omega_1,\omega_3) \end{bmatrix} E(\omega_1)|E(\omega_3)|^2 e^{-i\omega_1 t} \\ +\begin{bmatrix} \chi^{(3)}(\omega_2;-\omega_3,\omega_3,\omega_2) + \chi^{(3)}(\omega_2;\omega_3,-\omega_3,\omega_2) + \chi^{(3)}(\omega_2;\omega_2,\omega_3,-\omega_3) \\ +\chi^{(3)}(\omega_2;\omega_2,-\omega_3,\omega_3) + \chi^{(3)}(\omega_2;\omega_3,\omega_2,-\omega_3) + \chi^{(3)}(\omega_2;-\omega_3,\omega_2,\omega_3) \end{bmatrix} E(\omega_2)|E(\omega_3)|^2 e^{-i\omega_2 t} \end{Bmatrix} \\
& + c.c
\end{aligned} \tag{3.16}
$$

The combination of three fundamental frequencies produces 51 susceptibility tensors. First one is SDFG of $\omega_1$, $\omega_2$ and$\omega_3$, where the number of elements in the susceptibility tensors is 972 {[6(81+81(c.c.)) ($\omega_1+\omega_2-\omega_3$)]}. In DNFG process, number of elements in the susceptibility tensors is 1,458 {[3(81+81(c.c.)) ($\omega_1$)]+[3(81+81(c.c.))

Present address: Photonic Sciences Lab, Physical Research Laboratory, Navarangpura, Ahmedabad 380009, Gujarat, India, email: asvrao@prl.res.in

$\omega_2$)]+[3(81+81(c.c.)) $\omega_3$)]}.In NDNFG process, number of elements in the susceptibility tensors is 2,916 {[6(81+81(c.c.)) ($\omega_1$)]+[6(81+81(c.c.)) $\omega_2$)]+[6(81+81(c.c.)) $\omega_3$)]}. Therefore total number of elements is 5,346.In the off resonance the nonlinear polarization becomes

$$\begin{aligned}
\mathrm{P}^{(3)}(t) &= \frac{2}{3} d^{SDFG}_{\omega_1+\omega_2-\omega_3} \mathrm{E}(\omega_1)\mathrm{E}(\omega_2)\mathrm{E}^*(\omega_3) e^{-i(\omega_1+\omega_2-\omega_3)t} \\
&+ \frac{1}{3}\left\{ d^{DNGFG}_{\omega_1}\mathrm{E}(\omega_1)|\mathrm{E}(\omega_1)|^2 e^{-i\omega_1 t} + d^{DNGFG}_{\omega_2}\mathrm{E}(\omega_2)|\mathrm{E}(\omega_2)|^2 e^{-i\omega_2 t} + d^{DNGFG}_{\omega_3}\mathrm{E}(\omega_3)|\mathrm{E}(\omega_3)|^2 e^{-i\omega_3 t} \right\} \\
&+ \frac{2}{3}\left\{ \begin{array}{l} \left[ d^{NDNGFG}_{\omega_1}|\mathrm{E}(\omega_2)|^2 + d^{NDNGFG}_{\omega_1}|\mathrm{E}(\omega_3)|^2 \right]\mathrm{E}(\omega_1)e^{-i\omega_1 t} \\ + \left[ d^{NDNGFG}_{\omega_2}|\mathrm{E}(\omega_1)|^2 + d^{NDNGFG}_{\omega_2}|\mathrm{E}(\omega_3)|^2 \right]\mathrm{E}(\omega_2)e^{-i\omega_2 t} \\ + \left[ d^{NDNGFG}_{\omega_3}|\mathrm{E}(\omega_1)|^2 + d^{NDNGFG}_{\omega_3}|\mathrm{E}(\omega_2)|^2 \right]\mathrm{E}(\omega_3)e^{-i\omega_3 t} \end{array} \right\}
\end{aligned} \qquad (3.17)$$

The D-matrices will be same as we discussed in the THG. Total number of D-matrix elements is 300:{[30 (SDFG)]+[90 (DNFG)]+[180 (NDNFG)]}. In case of far off-resonance number of independent D-matrix elements is 150.

*3.4.SFG*

To produce SFG, let us fix the phase matching condition $\hat{k}_{\omega1+\omega2-\omega3}=\hat{k}_{\omega1}+\hat{k}_{\omega2}+\hat{k}_{\omega3}$ then from equation 3.2 the expression for nonlinear polarization is

$$\begin{aligned}
\mathrm{P}^{(3)}(t) &= \frac{1}{27}\begin{bmatrix} \chi^{(3)}(\omega_1+\omega_2+\omega_3;\omega_1,\omega_2,\omega_3) + \chi^{(3)}(\omega_1+\omega_3+\omega_2;\omega_1,\omega_3,\omega_2) \\ + \chi^{(3)}(\omega_2+\omega_1+\omega_3;\omega_2,\omega_1,\omega_3) + \chi^{(3)}(\omega_2+\omega_3+\omega_1;\omega_2,\omega_3,\omega_1) \\ + \chi^{(3)}(+\omega_3+\omega_1+\omega_2;\omega_3,\omega_1,\omega_2) + \chi^{(3)}(\omega_3+\omega_2+\omega_1;\omega_3,\omega_2,\omega_1) \end{bmatrix} \mathrm{E}(\omega_1)\mathrm{E}(\omega_2)\mathrm{E}(\omega_3)e^{-i(\omega_1+\omega_2+\omega_3)t} \\
&+ \frac{1}{27}\left\{ \begin{array}{l} \left[\chi^{(3)}(\omega_1;\omega_1,\omega_1,-\omega_1) + \chi^{(3)}(\omega_1;\omega_1,-\omega_1,\omega_1) + \chi^{(3)}(\omega_1;-\omega_1,\omega_1,\omega_1)\right]\mathrm{E}(\omega_1)|\mathrm{E}(\omega_1)|^2 e^{-i\omega_1 t} \\ + \left[\chi^{(3)}(\omega_2;\omega_2,\omega_2,-\omega_2) + \chi^{(3)}(\omega_2;\omega_2,-\omega_2,\omega_2) + \chi^{(3)}(\omega_2;-\omega_2,\omega_2,\omega_2)\right]\mathrm{E}(\omega_2)|\mathrm{E}(\omega_2)|^2 e^{-i\omega_2 t} \\ + \left[\chi^{(3)}(\omega_3;\omega_3,\omega_3,-\omega_3) + \chi^{(3)}(\omega_3;\omega_3,-\omega_3,\omega_3) + \chi^{(3)}(\omega_3;-\omega_3,\omega_3,\omega_3)\right]\mathrm{E}(\omega_3)|\mathrm{E}(\omega_3)|^2 e^{-i\omega_3 t} \end{array} \right\} \\
&+ \frac{1}{27}\left\{ \begin{array}{l}
+\begin{bmatrix} \chi^{(3)}(\omega_2;-\omega_1,\omega_1,\omega_2) + \chi^{(3)}(\omega_2;\omega_1,-\omega_1,\omega_2) + \chi^{(3)}(\omega_2;\omega_2,\omega_1,-\omega_1) \\ + \chi^{(3)}(\omega_2;\omega_2,-\omega_1,\omega_1) + \chi^{(3)}(\omega_2;\omega_1,\omega_2,-\omega_1) + \chi^{(3)}(\omega_2;-\omega_1,\omega_2,\omega_1) \end{bmatrix} \mathrm{E}(\omega_2)|\mathrm{E}(\omega_1)|^2 e^{-i\omega_2 t} \\
+\begin{bmatrix} \chi^{(3)}(\omega_3;-\omega_1,\omega_1,\omega_3) + \chi^{(3)}(\omega_3;\omega_1,-\omega_1,\omega_3) + \chi^{(3)}(\omega_3;\omega_3,\omega_1,-\omega_1) \\ + \chi^{(3)}(\omega_3;\omega_3,-\omega_1,\omega_1) + \chi^{(3)}(\omega_3;\omega_1,\omega_3,-\omega_1) + \chi^{(3)}(\omega_3;-\omega_1,\omega_3,\omega_1) \end{bmatrix} \mathrm{E}(\omega_1)|\mathrm{E}(\omega_1)|^2 e^{-i\omega_3 t} \\
+\begin{bmatrix} \chi^{(3)}(\omega_1;-\omega_2,\omega_2,\omega_1) + \chi^{(3)}(\omega_1;\omega_2,-\omega_2,\omega_1) + \chi^{(3)}(\omega_1;\omega_1,\omega_2,-\omega_2) \\ + \chi^{(3)}(\omega_1;\omega_1,-\omega_2,\omega_2) + \chi^{(3)}(\omega_1;\omega_2,\omega_1,-\omega_2) + \chi^{(3)}(\omega_1;-\omega_2,\omega_1,\omega_2) \end{bmatrix} \mathrm{E}(\omega_1)|\mathrm{E}(\omega_2)|^2 e^{-i\omega_1 t} \\
+\begin{bmatrix} \chi^{(3)}(\omega_3;-\omega_2,\omega_2,\omega_3) + \chi^{(3)}(\omega_3;\omega_2,-\omega_2,\omega_3) + \chi^{(3)}(\omega_3;\omega_3,\omega_2,-\omega_2) \\ + \chi^{(3)}(\omega_3;\omega_3,-\omega_2,\omega_2) + \chi^{(3)}(\omega_3;\omega_2,\omega_3,-\omega_2) + \chi^{(3)}(\omega_3;-\omega_2,\omega_3,\omega_2) \end{bmatrix} \mathrm{E}(\omega_3)|\mathrm{E}(\omega_2)|^2 e^{-i\omega_3 t} \\
+\begin{bmatrix} \chi^{(3)}(\omega_1;-\omega_3,\omega_3,\omega_1) + \chi^{(3)}(\omega_1;\omega_3,-\omega_3,\omega_1) + \chi^{(3)}(\omega_1;\omega_1,\omega_3,-\omega_3) \\ + \chi^{(3)}(\omega_1;\omega_1,-\omega_3,\omega_3) + \chi^{(3)}(\omega_1;\omega_3,\omega_1,-\omega_3) + \chi^{(3)}(\omega_1;-\omega_3,\omega_1,\omega_3) \end{bmatrix} \mathrm{E}(\omega_1)|\mathrm{E}(\omega_3)|^2 e^{-i\omega_1 t} \\
+\begin{bmatrix} \chi^{(3)}(\omega_2;-\omega_3,\omega_3,\omega_2) + \chi^{(3)}(\omega_2;\omega_3,-\omega_3,\omega_2) + \chi^{(3)}(\omega_2;\omega_2,\omega_3,-\omega_3) \\ + \chi^{(3)}(\omega_2;\omega_2,-\omega_3,\omega_3) + \chi^{(3)}(\omega_2;\omega_3,\omega_2,-\omega_3) + \chi^{(3)}(\omega_2;-\omega_3,\omega_2,\omega_3) \end{bmatrix} \mathrm{E}(\omega_2)|\mathrm{E}(\omega_3)|^2 e^{-i\omega_2 t}
\end{array} \right\} \\
&+ c.c
\end{aligned} \qquad (3.18)$$

The combination of three fundamental frequencies produces 51 susceptibility tensors. First one is SFG of $\omega_1$, $\omega_2$ and$\omega_3$, where the number of elements in the susceptibility tensors are 972 {[6(81+81(c.c.)) ($\omega_1+\omega_2+\omega_3$)]}. In DNFG process, number of elements in the susceptibility tensors is 1,458 {[3(81+81(c.c.)) ($\omega_1$)]+[3(81+81(c.c.))

$\omega_2$)]+[3(81+81(c.c.)) $\omega_3$)]}.In NDNFG process, number of elements in the susceptibility tensors is 2,916 {[6(81+81(c.c.)) ($\omega_1$)]+[6(81+81(c.c.)) $\omega_2$)]+[6(81+81(c.c.)) $\omega_3$)]}. Therefore total number of elements is 5,346. In case of off resonance the nonlinear polarization is

$$\begin{aligned}
\mathrm{P}^{(3)}(t) &= \frac{2}{3} d^{SFG}_{\omega_1+\omega_2+\omega_3} \mathrm{E}(\omega_1)\mathrm{E}(\omega_2)\mathrm{E}(\omega_3) e^{-i(\omega_1+\omega_2+\omega_3)t} \\
&+ \frac{1}{3}\left\{ d^{DNGFG}_{\omega_1} \mathrm{E}(\omega_1)\left|\mathrm{E}(\omega_1)\right|^2 e^{-i\omega_1 t} + d^{DNGFG}_{\omega_2} \mathrm{E}(\omega_2)\left|\mathrm{E}(\omega_2)\right|^2 e^{-i\omega_2 t} + d^{DNGFG}_{\omega_3} \mathrm{E}(\omega_3)\left|\mathrm{E}(\omega_3)\right|^2 e^{-i\omega_3 t} \right\} \\
&+ \frac{2}{3}\left\{ \begin{array}{l} \left[ d^{NDNGFG}_{\omega_1} \left|\mathrm{E}(\omega_2)\right|^2 + d^{NDNGFG}_{\omega_1} \left|\mathrm{E}(\omega_3)\right|^2 \right] \mathrm{E}(\omega_1) e^{-i\omega_1 t} \\ + \left[ d^{NDNGFG}_{\omega_2} \left|\mathrm{E}(\omega_1)\right|^2 + d^{NDNGFG}_{\omega_2} \left|\mathrm{E}(\omega_3)\right|^2 \right] \mathrm{E}(\omega_2) e^{-i\omega_2 t} \\ + \left[ d^{NDNGFG}_{\omega_3} \left|\mathrm{E}(\omega_1)\right|^2 + d^{NDNGFG}_{\omega_3} \left|\mathrm{E}(\omega_2)\right|^2 \right] \mathrm{E}(\omega_3) e^{-i\omega_3 t} \end{array} \right\}
\end{aligned} \tag{3.19}$$

Total number of D-matrix elements is 300:{[30 (THG)]+[90 (DNFG)]+[180 (NDNFG)]}. In case of far off-resonance number of independent D-matrix elements is 150.From equations 3.4, 3.17 and 3.19, the probability of generating THG is 9/2 larger than SFDFG and SFG of three waves.

## 4. Description of different nonlinear processes

After considerations have taken in the preceding sections 2 and 3 to second and third-order nonlinearity, in this section, the discussion is concerning light-matter interaction through considering their simultaneous presence in the nonlinear process. However, as shown in the previous sections, even though we are using phase matching conditions still other nonlinear terms are present in the required nonlinear process.

### *4.1. Harmonic generations*

Harmonic generations are parametric processes. Throughout this subsection, we have considered the far off-resonance process to reduce the complexity in the explanation of how different nonlinear phenomena can present simultaneously. In harmonic generations, NFG term generates the SPM in two ways: First one is Spatial Self-Phase Modulation (SSPM) and it will affect the transverse profile of beam as a consequence beam gets either self-focusing or defocusing. The second one is Temporal Self-Phase Modulation (TSPM) which can be seen in the pulsed laser. In pulsed lasers, the intensity variation along the propagation direction creates intensity dependent refractive index which modulates the temporal profile of the beam. Due to finite divergence of real beams, in the phase matching the k-vector will have the longitudinal and transverse components. The longitudinal component will be effected by TSPM and transverse components by SSPM. These effects come under third-order nonlinearity. In phase modulation there is another one, called as Mutual Phase Modulation (MPM), in this one wave produces nonlinear refractive index and other wave effects from this nonlinear refractive index. MPM always double than SPM because two frequencies are interacting in that process. In non-centrosymmetric crystals OR also affects the phase matching condition. OR of both fundamental and generated signals will be present. With propagation, intensity of fundamental frequency will be depleted and second harmonic signal generates. Therefore, as it propagates in the medium while nonlinear refractive index of OR due to fundamental beam decreases, due to harmonic generated intensity increase. Therefore OR and Phase modulation both contributes in the nonlinear refractive index. The phase shift due to nonlinear refractive index was first time observed in KTP waveguides by M. L. Sundheimer et.al in

Present address: Photonic Sciences Lab, Physical Research Laboratory, Navarangpura, Ahmedabad 380009, Gujarat, India, email: asvrao@prl.res.in

1993 [42]. Later it was demonstrated by Ch. Bosshard et.al in 1995 [43]. Thus in phase matching condition, it is necessary to include these intensity dependent terms to optimize the frequency conversion and to maintain the pulse parameters [44]. Depending upon the number of fundamental (incident) frequencies, different nonlinear processes will be generated. To simplify the susceptibility we reduce the $\chi_{ijk}(\omega_0+\omega_n+\omega_m+\ldots..;\omega_0, \omega_n, \omega_m, \ldots..)$ to $\chi_{ijk}(\omega_0+\omega_n+\omega_m+\ldots..)$. Eq.s from 4.1 to 4.8 give the expressions for nonlinear polarization for different harmonic generations. The terms in the first set are linear polarization of fundamental and harmonic frequencies. The second and third set terms correspond to second and third order nonlinear polarizations. In centrosymmetric crystal the second set can be neglected.

Nonlinear polarization for SHG

$$\begin{aligned}
\mathrm{P}(t) = &\left\{\chi^{(1)}(\omega)\mathrm{E}(\omega)e^{-i\omega t} + \chi^{(1)}(2\omega)\mathrm{E}(2\omega)e^{-i2\omega t}\right\} + 2\left\{d_{2\omega}^{\mathrm{SHG}}\mathrm{E}^2(\omega)e^{-i2\omega t} + d_{\omega}^{\mathrm{OR}}\left|\mathrm{E}(\omega)\right|^2 + d_{2\omega}^{\mathrm{OR}}\left|\mathrm{E}(2\omega)\right|^2\right\} \\
&+ 9\left\{\begin{aligned} &\mathrm{d}_{\omega}^{\mathrm{SPM}}\mathrm{E}(\omega)\left|\mathrm{E}(\omega)\right|^2 e^{-i\omega t} + \mathrm{d}_{2\omega}^{\mathrm{SPM}}\mathrm{E}(2\omega)\left|\mathrm{E}(2\omega)\right|^2 e^{-i2\omega t} \\ &+ 2\mathrm{d}_{\omega}^{\mathrm{MPM}}\mathrm{E}(\omega)\left|\mathrm{E}(2\omega)\right|^2 e^{-i\omega t} + 2\mathrm{d}_{2\omega}^{\mathrm{SPM}}\mathrm{E}(2\omega)\left|\mathrm{E}(\omega)\right|^2 e^{-i2\omega t}\end{aligned}\right\}
\end{aligned} \tag{4.1}$$

Nonlinear polarization for THG

$$\begin{aligned}
\mathrm{P}(t) = &\left\{\chi^{(1)}(\omega)\mathrm{E}(\omega)e^{-i\omega t} + \chi^{(1)}(3\omega)\mathrm{E}(3\omega)e^{-i3\omega t}\right\} + 2\left\{d_{\omega}^{\mathrm{OR}}\left|\mathrm{E}(\omega)\right|^2 + d_{3\omega}^{\mathrm{OR}}\left|\mathrm{E}(3\omega)\right|^2\right\} \\
&+ 9\left\{\begin{aligned} &\frac{1}{3}d_{3\omega}^{\mathrm{THG}}\mathrm{E}^3(\omega)e^{-i3\omega t} + d_{\omega}^{\mathrm{SPM}}\mathrm{E}(\omega)\left|\mathrm{E}(\omega)\right|^2 \mathrm{E}e^{-i\omega t} + d_{3\omega}^{\mathrm{SPM}}\left|\mathrm{E}(3\omega)\right|^2 \mathrm{E}e^{-i3\omega t} \\ &+ 2d_{\omega}^{\mathrm{MPM}}\mathrm{E}(\omega)\left|\mathrm{E}(3\omega)\right|^2 \mathrm{E}e^{-i\omega t} + 2d_{3\omega}^{\mathrm{MPM}}\mathrm{E}(3\omega)\left|\mathrm{E}(\omega)\right|^2 \mathrm{E}e^{-i3\omega t}\end{aligned}\right\}
\end{aligned} \tag{4.2}$$

Nonlinear polarization for SFG

$$\begin{aligned}
\mathrm{P}(t) = &\left\{\frac{1}{2}\chi^{(1)}(\omega_1)\mathrm{E}(\omega_1)e^{-i\omega_1 t} + \frac{1}{2}\chi^{(1)}(\omega_2)\mathrm{E}(\omega_2)e^{-i\omega_2 t} + \chi^{(1)}(\omega_1+\omega_2)\mathrm{E}(\omega_1+\omega_2)e^{-i(\omega_1+\omega_2)t}\right\} \\
&+ \left\{d_{\omega_1+\omega_2}^{\mathrm{SHG}}\mathrm{E}(\omega_1)\mathrm{E}(\omega_2)e^{-i(\omega_1+\omega_2)t} + d_{\omega_1}^{\mathrm{OR}}\left|\mathrm{E}(\omega_1)\right|^2 + d_{\omega_2}^{\mathrm{OR}}\left|\mathrm{E}(\omega_2)\right|^2 + 2d_{\omega_1+\omega_2}^{\mathrm{OR}}\left|\mathrm{E}(\omega_1+\omega_2)\right|^2\right\} \\
&+ 9\left\{\begin{aligned} &\left[\frac{1}{8}d_{\omega_1}^{\mathrm{SPM}}\mathrm{E}(\omega_1)\left|\mathrm{E}(\omega_1)\right|^2 + \frac{1}{4}d_{\omega_1}^{\mathrm{MPM}}\mathrm{E}(\omega_1)\left|\mathrm{E}(\omega_2)\right|^2 + d_{\omega_1}^{\mathrm{MPM}}\mathrm{E}(\omega_1)\left|\mathrm{E}(\omega_1+\omega_2)\right|^2\right]e^{-i\omega_1 t} \\ &+ \left[\frac{1}{8}d_{\omega_2}^{\mathrm{SPM}}\mathrm{E}(\omega_2)\left|\mathrm{E}(\omega_2)\right|^2 + \frac{1}{4}d_{\omega_2}^{\mathrm{MPM}}\mathrm{E}(\omega_2)\left|\mathrm{E}(\omega_1)\right|^2 + d_{\omega_2}^{\mathrm{MPM}}\mathrm{E}(\omega_2)\left|\mathrm{E}(\omega_1+\omega_2)\right|^2\right]e^{-i\omega_2 t} \\ &+ \left[\begin{aligned} &d_{\omega_1+\omega_2}^{\mathrm{SPM}}\left|\mathrm{E}(\omega_1+\omega_2)\right|^2 \mathrm{E}(\omega_1+\omega_2)e^{-i(\omega_1+\omega_2)t} \\ &+ \frac{1}{2}d_{\omega_1+\omega_2}^{\mathrm{MPM}}\left|\mathrm{E}(\omega_1)\right|^2 \mathrm{E}(\omega_1+\omega_2)e^{-i(\omega_1+\omega_2)t} + \frac{1}{2}d_{\omega_1+\omega_2}^{\mathrm{MPM}}\left|\mathrm{E}(\omega_2)\right|^2 \mathrm{E}(\omega_1+\omega_2)e^{-i(\omega_1+\omega_2)t}\end{aligned}\right]\end{aligned}\right\}
\end{aligned} \tag{4.3}$$

Nonlinear polarization for DFG

$$
\begin{aligned}
\mathrm{P}(t) = &\left\{ \frac{1}{2}\chi^{(1)}(\omega_1)\mathrm{E}(\omega_1)e^{-i\omega_1 t} + \frac{1}{2}\chi^{(1)}(\omega_2)\mathrm{E}(\omega_2)e^{-i\omega_2 t} + \chi^{(1)}(\omega_2-\omega_1)\mathrm{E}(\omega_2-\omega_1)e^{-i(\omega_2-\omega_1)t} \right\} \\
&+ \left\{ d^{\mathrm{DFG}}_{\omega_2-\omega_1}\mathrm{E}^*(\omega_1)\mathrm{E}(\omega_2)\left\langle e^{-i\omega_1 t} \middle| e^{-i(\omega_2-\omega_1)t} \right\rangle + d^{\mathrm{OR}}_{\omega_1}\left|\mathrm{E}(\omega_1)\right|^2 + d^{\mathrm{OR}}_{\omega_2}\left|\mathrm{E}(\omega_2)\right|^2 + 2d^{\mathrm{OR}}_{\omega_2-\omega_1}\left|\mathrm{E}(\omega_2-\omega_1)\right|^2 \right\} \\
&+ 9\left\{ \begin{aligned}
&\left[ \frac{1}{8}d^{\mathrm{SPM}}_{\omega_1}\mathrm{E}(\omega_1)\left|\mathrm{E}(\omega_1)\right|^2 + \frac{1}{4}d^{\mathrm{MPM}}_{\omega_1}\mathrm{E}(\omega_1)\left|\mathrm{E}(\omega_2)\right|^2 + d^{\mathrm{MPM}}_{\omega_1}\mathrm{E}(\omega_1)\left|\mathrm{E}(\omega_2-\omega_1)\right|^2 \right] e^{-i\omega_1 t} \\
&+ \left[ \frac{1}{8}d^{\mathrm{SPM}}_{\omega_2}\mathrm{E}(\omega_2)\left|\mathrm{E}(\omega_2)\right|^2 + \frac{1}{4}d^{\mathrm{MPM}}_{\omega_2}\mathrm{E}(\omega_2)\left|\mathrm{E}(\omega_1)\right|^2 + d^{\mathrm{MPM}}_{\omega_2}\mathrm{E}(\omega_2)\left|\mathrm{E}(\omega_2-\omega_1)\right|^2 \right] e^{-i\omega_2 t} \\
&+ \left[ \begin{aligned}
&\mathrm{d}^{\mathrm{SPM}}_{\omega_2-\omega_1}\left|\mathrm{E}(\omega_2-\omega_1)\right|^2\mathrm{E}(\omega_1-\omega_2)e^{-i(\omega_2-\omega_1)t} \\
&+ \frac{1}{2}\mathrm{d}^{\mathrm{MPM}}_{\omega_2-\omega_1}\left|\mathrm{E}(\omega_1)\right|^2\mathrm{E}(\omega_1-\omega_2)e^{-i(\omega_2-\omega_1)t} + \frac{1}{2}\mathrm{d}^{\mathrm{MPM}}_{\omega_2-\omega_1}\left|\mathrm{E}(\omega_2)\right|^2\mathrm{E}(\omega_1-\omega_2)e^{-i(\omega_2-\omega_1)t}
\end{aligned} \right]
\end{aligned} \right\}
\end{aligned}
\tag{4.4}
$$

Nonlinear polarization for SFSHG

$$
\begin{aligned}
\mathrm{P}(t) = &\left\{ \frac{1}{2}\chi^{(1)}(\omega_1)\mathrm{E}(\omega_1)e^{-i\omega_1 t} + \frac{1}{2}\chi^{(1)}(\omega_2)\mathrm{E}(\omega_2)e^{-i\omega_2 t} + \chi^{(1)}(2\omega_1+\omega_2)\mathrm{E}(2\omega_1+\omega_2)e^{-i(2\omega_1+\omega_2)t} \right\} \\
&+ \left\{ d^{\mathrm{OR}}_{\omega_1}\left|\mathrm{E}(\omega_1)\right|^2 + d^{\mathrm{OR}}_{\omega_2}\left|\mathrm{E}(\omega_2)\right|^2 + 2d^{\mathrm{OR}}_{2\omega_1+\omega_2}\left|\mathrm{E}(2\omega_1+\omega_2)\right|^2 \right\} \\
&+ 9\left\{ \begin{aligned}
&\frac{1}{8}d^{\mathrm{SFSHG}}_{2\omega_1+\omega_2}\mathrm{E}^2(\omega_1)\mathrm{E}(\omega_2)e^{-i(2\omega_1+\omega_2)t} \\
&+ \left[ \frac{1}{8}d^{\mathrm{SPM}}_{\omega_1}\mathrm{E}_1(\omega_1)\left|\mathrm{E}(\omega_1)\right|^2 + \frac{1}{4}d^{\mathrm{MPM}}_{\omega_1}\mathrm{E}(\omega_1)\left|\mathrm{E}(\omega_2)\right|^2 + d^{\mathrm{MPM}}_{\omega_1}\mathrm{E}(\omega_1)\left|\mathrm{E}(2\omega_1+\omega_2)\right|^2 \right] e^{-i\omega_1 t} \\
&+ \left[ \frac{1}{8}d^{\mathrm{SPM}}_{\omega_2}\mathrm{E}(\omega_2)\left|\mathrm{E}(\omega_2)\right|^2 + \frac{1}{4}d^{\mathrm{MPM}}_{\omega_2}\mathrm{E}(\omega_2)\left|\mathrm{E}(\omega_1)\right|^2 + d^{\mathrm{MPM}}_{\omega_2}\mathrm{E}(\omega_2)\left|\mathrm{E}(2\omega_1+\omega_2)\right|^2 \right] e^{-i\omega_2 t} \\
&+ \left\{ \begin{aligned}
&\mathrm{d}^{\mathrm{SPM}}_{2\omega_1+\omega_2}\mathrm{E}(2\omega_1+\omega_2)\left|\mathrm{E}(2\omega_1+\omega_2)\right|^2 e^{-i(2\omega_1+\omega_2)t} \\
&+ \frac{1}{2}\mathrm{d}^{\mathrm{MPM}}_{2\omega_1+\omega_2}\mathrm{E}(2\omega_1+\omega_2)\left|\mathrm{E}(\omega_1)\right|^2 e^{-i(2\omega_1+\omega_2)t} + \frac{1}{2}\mathrm{d}^{\mathrm{MPM}}_{2\omega_1+\omega_2}\mathrm{E}(2\omega_1+\omega_2)\left|\mathrm{E}(\omega_2)\right|^2 e^{-i(2\omega_1+\omega_2)t}
\end{aligned} \right\}
\end{aligned} \right\}
\end{aligned}
\tag{4.5}
$$

Nonlinear polarization for DFSHG

$$
\begin{aligned}
\mathrm{P}(t) = &\left\{ \frac{1}{2}\chi^{(1)}(\omega_1)\mathrm{E}(\omega_1)e^{-i\omega_1 t} + \frac{1}{2}\chi^{(1)}(\omega_2)\mathrm{E}(\omega_2)e^{-i\omega_2 t} + \frac{1}{2}\chi^{(1)}(2\omega_1-\omega_2)\mathrm{E}(2\omega_1-\omega_2)e^{-i(2\omega_1-\omega_2)t} \right\} \\
&+ \left\{ d^{\mathrm{OR}}_{11}\left|\mathrm{E}(\omega_1)\right|^2 + d^{\mathrm{OR}}_{11}\left|\mathrm{E}(\omega_2)\right|^2 + 2d^{\mathrm{OR}}_{2\omega_1-\omega_2}\left|\mathrm{E}(2\omega_1-\omega_2)\right|^2 \right\} \\
&+ 9\left\{ \begin{aligned}
&\frac{1}{8}d^{\mathrm{DFSHG}}_{2\omega_1-\omega_2}\mathrm{E}^2(\omega_1)\mathrm{E}(\omega_2)\left\langle e^{-i\omega_2 t} \middle| e^{-i(2\omega_1-\omega_2)t} \right\rangle \\
&+ \left[ \frac{1}{8}d^{\mathrm{SPM}}_{\omega_1}\mathrm{E}(\omega_1)\left|\mathrm{E}(\omega_1)\right|^2 + \frac{1}{4}d^{\mathrm{MPM}}_{\omega_1}\mathrm{E}(\omega_1)\left|\mathrm{E}(\omega_2)\right|^2 + d^{\mathrm{MPM}}_{\omega_1}\mathrm{E}(\omega_1)\left|\mathrm{E}(2\omega_1-\omega_2)\right|^2 \right] e^{-i\omega_1 t} \\
&+ \left[ \frac{1}{8}d^{\mathrm{SPM}}_{\omega_2}\mathrm{E}(\omega_2)\left|\mathrm{E}(\omega_2)\right|^2 + \frac{1}{4}d^{\mathrm{MPM}}_{\omega_2}\mathrm{E}(\omega_2)\left|\mathrm{E}(\omega_1)\right|^2 + d^{\mathrm{MPM}}_{\omega_2}\mathrm{E}(\omega_2)\left|\mathrm{E}(2\omega_1-\omega_2)\right|^2 \right] e^{-i\omega_2 t} \\
&+ \left[ \begin{aligned}
&\mathrm{d}^{\mathrm{SPM}}_{2\omega_1-\omega_2}\mathrm{E}(2\omega_1-\omega_2)\left|\mathrm{E}(2\omega_1-\omega_2)\right|^2 e^{-i(2\omega_1-\omega_2)t} \\
&\frac{1}{2}\mathrm{d}^{\mathrm{SPM}}_{2\omega_1-\omega_2}\mathrm{E}(2\omega_1-\omega_2)\left|\mathrm{E}(\omega_1)\right|^2 e^{-i(2\omega_1-\omega_2)t} + \frac{1}{2}\mathrm{d}^{\mathrm{SPM}}_{2\omega_1-\omega_2}\mathrm{E}(2\omega_1-\omega_2)\left|\mathrm{E}(\omega_2)\right|^2 e^{-i(2\omega_1-\omega_2)t}
\end{aligned} \right]
\end{aligned} \right\}
\end{aligned}
\tag{4.6}
$$

Nonlinear polarization for SFDFG

Present address: Photonic Sciences Lab, Physical Research Laboratory, Navarangpura, Ahmedabad 380009, Gujarat, India, email: asvrao@prl.res.in

$$\mathrm{P}(t)=\left\{\frac{1}{3}\chi^{(1)}(\omega_1)\mathrm{E}(\omega_1)e^{-i\omega_1 t}+\frac{1}{3}\chi^{(1)}(\omega_2)\mathrm{E}(\omega_2)e^{-i\omega_2 t}+\frac{1}{3}\chi^{(1)}(\omega_3)\mathrm{E}(\omega_3)e^{-i\omega_3 t}+\chi^{(1)}(\omega_1+\omega_2-\omega_3)\mathrm{E}(\omega_1+\omega_2-\omega_3)e^{-i(\omega_1+\omega_2-\omega_3)t}\right\}$$
$$+\left\{\frac{2}{9}d^{\mathrm{OR}}_{\omega_1}|\mathrm{E}(\omega_1)|^2+\frac{2}{9}d^{\mathrm{OR}}_{\omega_2}|\mathrm{E}(\omega_2)|^2+\frac{2}{9}d^{\mathrm{OR}}_{\omega_3}|\mathrm{E}(\omega_3)|^2+2d^{\mathrm{OR}}_{\omega_1+\omega_2-\omega_3}|\mathrm{E}(\omega_1+\omega_2-\omega_3)|^2\right\}$$
$$+\left\{\begin{array}{l}\frac{2}{3}d^{\mathrm{DSFG}}_{\omega_1+\omega_2-\omega_3}\mathrm{E}(\omega_1)\mathrm{E}(\omega_2)\mathrm{E}^*(\omega_3)\left\langle e^{-i\omega_3 t}\middle|e^{-i(\omega_1+\omega_2-\omega_3)t}\right\rangle\\
+\left[\frac{1}{3}d^{\mathrm{SPM}}_{\omega_1}\mathrm{E}(\omega_1)|\mathrm{E}(\omega_1)|^2+\frac{2}{3}d^{\mathrm{MPM}}_{\omega_1}\mathrm{E}(\omega_1)|\mathrm{E}(\omega_2)|^2+\frac{2}{3}d^{\mathrm{MPM}}_{\omega_1}\mathrm{E}(\omega_1)|\mathrm{E}(\omega_3)|^2+6d^{\mathrm{MPM}}_{\omega_1}\mathrm{E}(\omega_1)|\mathrm{E}(\omega_1+\omega_2-\omega_3)|^2\right]e^{-i\omega_1 t}\\
+\left[\frac{1}{3}d^{\mathrm{SPM}}_{\omega_2}\mathrm{E}(\omega_2)|\mathrm{E}(\omega_2)|^2+\frac{2}{3}d^{\mathrm{MPM}}_{\omega_2}\mathrm{E}(\omega_2)|\mathrm{E}(\omega_1)|^2+\frac{2}{3}d^{\mathrm{MPM}}_{\omega_2}\mathrm{E}(\omega_2)|\mathrm{E}(\omega_3)|^2+6d^{\mathrm{MPM}}_{\omega_2}\mathrm{E}(\omega_2)|\mathrm{E}(\omega_1+\omega_2-\omega_3)|^2\right]e^{-i\omega_2 t}\\
+\left[\frac{1}{3}d^{\mathrm{SPM}}_{\omega_3}\mathrm{E}(\omega_3)|\mathrm{E}(\omega_3)|^2+\frac{2}{3}d^{\mathrm{MPM}}_{\omega_2}\mathrm{E}(\omega_3)|\mathrm{E}(\omega_1)|^2+\frac{2}{3}d^{\mathrm{MPM}}_{\omega_2}\mathrm{E}(\omega_3)|\mathrm{E}(\omega_2)|^2+6d^{\mathrm{MPM}}_{\omega_3}\mathrm{E}(\omega_3)|\mathrm{E}(\omega_1+\omega_2-\omega_3)|^2\right]e^{-i\omega_3 t}\\
+\left[\begin{array}{l}9\mathrm{d}^{\mathrm{SPM}}_{\omega_1+\omega_2-\omega_3}\mathrm{E}(\omega_1+\omega_2-\omega_3)|\mathrm{E}(\omega_1+\omega_2-\omega_3)|^2e^{-i(\omega_1+\omega_2-\omega_3)t}+2\mathrm{d}^{\mathrm{MPM}}_{\omega_1+\omega_2-\omega_3}\mathrm{E}(\omega_1+\omega_2-\omega_3)|\mathrm{E}(\omega_1)|^2e^{-i(\omega_1+\omega_2-\omega_3)t}\\+2\mathrm{d}^{\mathrm{MPM}}_{\omega_1+\omega_2-\omega_3}\mathrm{E}(\omega_1+\omega_2-\omega_3)|\mathrm{E}(\omega_2)|^2e^{-i(\omega_1+\omega_2-\omega_3)t}+2\mathrm{d}^{\mathrm{MPM}}_{\omega_1+\omega_2-\omega_3}\mathrm{E}(\omega_1+\omega_2-\omega_3)|\mathrm{E}(\omega_3)|^2e^{-i(\omega_1+\omega_2-\omega_3)t}\end{array}\right]\end{array}\right\} \quad (4.7)$$

Nonlinear polarization for SFG (three frequencies)

$$\mathrm{P}(t)=\left\{\frac{1}{3}\chi^{(1)}(\omega_1)\mathrm{E}(\omega_1)e^{-i\omega_1 t}+\frac{1}{3}\chi^{(1)}(\omega_2)\mathrm{E}(\omega_2)e^{-i\omega_2 t}+\frac{1}{3}\chi^{(1)}(\omega_3)\mathrm{E}(\omega_3)e^{-i\omega_3 t}+\chi^{(1)}(\omega_1+\omega_2+\omega)\mathrm{E}(\omega_1+\omega_2+\omega)e^{-i(\omega_1+\omega_2+\omega)t}\right\}$$
$$+\left\{\frac{2}{9}d^{\mathrm{OR}}_{\omega_1}|\mathrm{E}(\omega_1)|^2+\frac{2}{9}d^{\mathrm{OR}}_{\omega_2}|\mathrm{E}(\omega_2)|^2+\frac{2}{9}d^{\mathrm{OR}}_{\omega_3}|\mathrm{E}(\omega_3)|^2+2d^{\mathrm{OR}}_{\omega_1+\omega_2+\omega}|\mathrm{E}(\omega_1+\omega_2+\omega)|^2\right\}$$
$$+\left\{\begin{array}{l}\frac{2}{3}d^{\mathrm{SFG}}_{\omega_1+\omega_2+\omega_3}\mathrm{E}(\omega_1)\mathrm{E}(\omega_2)\mathrm{E}_1(\omega_2)e^{-i(\omega_1+\omega_2+\omega_3)t}\\
+\left[\frac{1}{3}d^{\mathrm{SPM}}_{\omega_1}\mathrm{E}(\omega_1)|\mathrm{E}_1(\omega_1)|^2+\frac{2}{3}d^{\mathrm{MPM}}_{\omega_1}\mathrm{E}(\omega_1)|\mathrm{E}(\omega_2)|^2+\frac{2}{3}d^{\mathrm{MPM}}_{\omega_1}\mathrm{E}(\omega_1)|\mathrm{E}(\omega_3)|^2+6d^{\mathrm{MPM}}_{\omega_1}\mathrm{E}(\omega_1)|\mathrm{E}(\omega_1+\omega_2+\omega_3)|^2\right]e^{-i\omega_1 t}\\
+\left[\frac{1}{3}d^{\mathrm{SPM}}_{\omega_2}\mathrm{E}(\omega_2)|\mathrm{E}(\omega_2)|^2+\frac{2}{3}d^{\mathrm{MPM}}_{\omega_2}\mathrm{E}(\omega_2)|\mathrm{E}(\omega_1)|^2+\frac{2}{3}d^{\mathrm{MPM}}_{\omega_2}\mathrm{E}(\omega_2)|\mathrm{E}(\omega_3)|^2+6d^{\mathrm{MPM}}_{\omega_2}\mathrm{E}(\omega_2)|\mathrm{E}(\omega_1+\omega_2+\omega_3)|^2\right]e^{-i\omega_2 t}\\
+\left[\frac{1}{3}d^{\mathrm{SPM}}_{\omega_3}\mathrm{E}(\omega_3)|\mathrm{E}(\omega_3)|^2+\frac{2}{3}d^{\mathrm{MPM}}_{\omega_3}\mathrm{E}(\omega_3)|\mathrm{E}(\omega_1)|^2+\frac{2}{3}d^{\mathrm{MPM}}_{\omega_3}\mathrm{E}(\omega_3)|\mathrm{E}(\omega_2)|^2+6d^{\mathrm{MPM}}_{\omega_3}\mathrm{E}(\omega_3)|\mathrm{E}(\omega_1+\omega_2+\omega_3)|^2\right]e^{-i\omega_3 t}\\
+\left[\begin{array}{l}9\mathrm{d}^{\mathrm{SPM}}_{\omega_1+\omega_2+\omega_3}\mathrm{E}(\omega_1+\omega_2+\omega_3)|\mathrm{E}(\omega_1+\omega_2+\omega_3)|^2e^{-i(\omega_1+\omega_2+\omega_3)t}+2\mathrm{d}^{\mathrm{SPM}}_{\omega_1+\omega_2+\omega_3}\mathrm{E}(\omega_1+\omega_2+\omega_3)|\mathrm{E}(\omega_1)|^2e^{-i(\omega_1+\omega_2+\omega_3)t}\\+2\mathrm{d}^{\mathrm{SPM}}_{\omega_1+\omega_2+\omega_3}\mathrm{E}(\omega_1+\omega_2+\omega_3)|\mathrm{E}(\omega_2)|^2e^{-i(\omega_1+\omega_2+\omega_3)t}+2\mathrm{d}^{\mathrm{SPM}}_{\omega_1+\omega_2+\omega_3}\mathrm{E}(\omega_1+\omega_2+\omega_3)|\mathrm{E}(\omega_3)|^2e^{-i(\omega_1+\omega_2+\omega_3)t}\end{array}\right]\end{array}\right\} \quad (4.8)$$

### *4.2. Non Harmonic generations*

Non harmonic generations come under nonparametric processes. To explain these processes let us consider two optical amplitudes coupling under third order nonlinear process then the nonlinear polarization is

$$\mathrm{P}^{(3)}(t)=\frac{1}{8}\left\{\begin{array}{l}\left[\chi^{(3)}(\omega_1;\omega_1,\omega_1,-\omega_1)+\chi^{(3)}(\omega_1;\omega_1,-\omega_1,\omega_1)+\chi^{(3)}(\omega_1;-\omega_1,\omega_1,\omega_1)\right]\mathrm{E}(\omega_1)|\mathrm{E}(\omega_1)|^2e^{-i\omega_1 t}\\+\left[\chi^{(3)}(\omega_2;\omega_2,\omega_2,-\omega_2)+\chi^{(3)}(\omega_2;\omega_2,-\omega_2,\omega_2)+\chi^{(3)}(\omega_2;-\omega_2,\omega_2,\omega_2)\right]\mathrm{E}(\omega_2)|\mathrm{E}(\omega_2)|^2e^{-i\omega_2 t}\end{array}\right\}$$
$$+\frac{1}{8}\left\{\begin{array}{l}+\left[\begin{array}{l}\chi^{(3)}(\omega_2;-\omega_1,\omega_1,\omega_2)+\chi^{(3)}(\omega_2;\omega_1,-\omega_1,\omega_2)+\chi^{(3)}(\omega_2;\omega_2,\omega_1,-\omega_1)\\+\chi^{(3)}(\omega_2;\omega_2,-\omega_1,\omega_1)+\chi^{(3)}(\omega_2;\omega_1,\omega_2,-\omega_1)+\chi^{(3)}(\omega_2;-\omega_1,\omega_2,\omega_1)\end{array}\right]\mathrm{E}(\omega_2)|\mathrm{E}(\omega_1)|^2e^{-i\omega_2 t}\\
+\left[\begin{array}{l}\chi^{(3)}(\omega_1;-\omega_2,\omega_2,\omega_1)+\chi^{(3)}(\omega_1;\omega_2,-\omega_2,\omega_1)+\chi^{(3)}(\omega_1;\omega_1,\omega_2,-\omega_2)\\+\chi^{(3)}(\omega_1;\omega_1,-\omega_2,\omega_2)+\chi^{(3)}(\omega_1;\omega_2,\omega_1,-\omega_2)+\chi^{(3)}(\omega_1;-\omega_2,\omega_1,\omega_2)\end{array}\right]\mathrm{E}(\omega_1)|\mathrm{E}(\omega_2)|^2e^{-i\omega_1 t}\end{array}\right\}$$
$$+c.c. \quad (4.9)$$

The terms in the first set of Eq. 4.9 give the degenerative nonlinear process, where single frequency present in the nonlinear process. The imaginary part corresponds to nonlinear absorptive index (nonlinear absorption) and real part

Present address: Photonic Sciences Lab, Physical Research Laboratory, Navarangpura, Ahmedabad 380009, Gujarat, India, email: asvrao@prl.res.in

corresponds to the nonlinear refractive index. These terms can be measured by nonlinear techniques like single beam transmittance [43-45] and degenerative nonlinear wave mixing [46-48]. This nonlinearity can be seen in the physical process like Absorption, Molecular orientation, Polarization etc.  The second set of terms gives the nonlinearity under non-degenerative fundamental frequencies. The processes like non-degenerative absorption, Stimulated Raman and Brillion scattering comes under this category. These nonparametric processes widely used in spectroscopic analysis and form a branch called as nonlinear spectroscopy.

## 5. Conclusion

In this article, we have explicitly elaborated the nonlinear polarization of second and third order in terms of susceptibility tensors. In three cases: resonant, off-resonant and far off-resonant; I have calculated the number of susceptibility terms present in the nonlinear polarization and have shown how the susceptibility tensor is reduced while we are going from resonant case to far off-resonant case. For a given incident intensity the probability of generating SHG is two times greater than SFG and DFG under the exception of material activeness in their generation. Also, DFG is little less probable due to its requirement of stimulated down conversion as compare with SHG and SFG. I have shown the effect of the nonlinear refractive index which is present in the harmonic generation caused by OR and third-order nonlinear refractive index on the phase matching condition for different harmonic generations. The probabilities of generating the third order harmonics are THG: (SFSHG, DFSHG): (SFG, SFDFG)=1:3/8:2/9 but physically DSHG and SFDFG are less probable because in their process stimulated down conversion is needed. I have shown the origin of nonlinear phase modulation and how it is so imperative in phase matching of harmonic generation. At the end of the discussion, I have given a short note on nonparametric processes basically differ from Harmonic generation. This article not concerned about depletion in the beams and effect of materials on the probabilities of each nonlinear process, because of detailed discussion of these effects already discussed in the references [49-55].

Present address: Photonic Sciences Lab, Physical Research Laboratory, Navarangpura, Ahmedabad 380009, Gujarat, India, email: asvrao@prl.res.in

Present address: Photonic Sciences Lab, Physical Research Laboratory, Navarangpura, Ahmedabad 380009, Gujarat, India, email: asvrao@prl.res.in

Present address: Photonic Sciences Lab, Physical Research Laboratory, Navarangpura, Ahmedabad 380009, Gujarat, India, email: asvrao@prl.res.in

Present address: Photonic Sciences Lab, Physical Research Laboratory, Navarangpura, Ahmedabad 380009, Gujarat, India, email: asvrao@prl.res.in

Present address: Photonic Sciences Lab, Physical Research Laboratory, Navarangpura, Ahmedabad 380009, Gujarat, India, email: asvrao@prl.res.in